\documentclass[lettersize,journal]{IEEEtran}

\usepackage{amsmath,amsfonts}
\usepackage[ruled,vlined]{algorithm2e}
\usepackage{array}
\usepackage{textcomp}
\usepackage{stfloats}
\usepackage{url}
\usepackage{subfigure}
\usepackage{verbatim}
\usepackage{graphicx}
\usepackage{cite,color}
\usepackage{stmaryrd}
\usepackage{caption}
\usepackage{ragged2e}
\usepackage{scalerel}
\usepackage{graphicx}
\usepackage{amsmath,amssymb,amsfonts}
\makeatletter
\renewcommand{\maketag@@@}[1]{\hbox{\m@th\normalsize\normalfont#1}}%
\makeatother
\usepackage{mathrsfs}
\usepackage{amsthm}
\usepackage{xcolor}
\definecolor{MXY}{RGB}{252,8,235}
\usepackage[font={small,it}]{caption}

\usepackage{booktabs}
\usepackage{adjustbox}
\usepackage{xcolor}

\begin{document}
\newtheorem{lemma}{Lemma}
\newtheorem{theorem}{Theorem}
\newtheorem{corollary}{Corollary}
\title{Transmit Coefficients and Receive Combining\\ Vector Design for OTA-FL with Imperfect CSI}

\author{Xiaoyan Ma,~\IEEEmembership{Member,~IEEE}, Shahryar Zehtabi,~\IEEEmembership{Student Member,~IEEE}, Yinan Zou,~\IEEEmembership{Student Member,~IEEE}, Taejoon Kim,~\IEEEmembership{Senior Member,~IEEE}
and Christopher G. Brinton,~\IEEEmembership{Senior Member,~IEEE}
\thanks{Xiaoyan Ma, Shahryar Zehtabi, Yinan Zou and Christopher G. Brinton are from School of Electrical and Computer Engineering, Purdue University, Email: \{ma946, szehtabi, zou211, cgb\}@purdue.edu. Taejoon Kim is from School of Electrical, Computer and Energy Engineering, Arizona State University,  Email: taejoonkim@asu.edu. This work was supported in part by the National Science Foundation (NSF) under grants CNS-2146171, CNS-2451268,
CNS2514415, ITE251537, and ET-2554467, and by the Office of Naval Research (ONR) under grant N00014-21-1-2472.}
}

\maketitle

\begin{abstract} 
Over-the-air (OTA) computation has recently gained significant attentions as an effective approach to enhance the communication efficiency of wireless federated learning (FL). By enabling simultaneous transmission and aggregation of local model updates, OTA-FL can substantially reduce both latency and bandwidth consumption.
However, a key challenge lies in the imperfect aggregation of global models caused by channel state information (CSI) uncertainty, which introduces distortion to the final learning performance. 
To address this issue, we study the long-term mean squared error (MSE) minimization problem for OTA-FL under imperfect CSI conditions. 
Through convergence analysis, we establish an upper bound for the time-averaged MSE, thereby revealing the effect of aggregation errors accumulated throughout multiple communication rounds on the overall training performances. 
Based on this analysis, an optimization framework is developed to minimize the long-term MSE via the joint design of (i) transmit coefficients at the local devices and (ii) receive combining vectors at the parameter server (PS). 
Since this alternating optimization approach requires non-causal CSI, a Lyapunov-based optimization method is further introduced to handle causal CSI scenarios. 
By incorporating virtual queues to characterize long-term energy consumption, the proposed method effectively decouples temporal dependencies and allows transmit coefficients to be optimized based on the causal CSI of each aggregation round. 
Comprehensive evaluations on Fashion-MNIST, CIFAR-10 and CIFAR-100 datasets have demonstrated that the proposed algorithms can significantly reduce the degradation of test accuracy caused by imperfect CSI. Comparisons with other benchmark schemes further verify the superiority of our proposed algorithms.


\end{abstract}


\section{Introduction}
With the rapid proliferation of wireless Internet of Things (IoT) devices, 
a massive amount of data is continuously produced and retained at distributed local devices \cite{cgb}. Traditional machine learning (ML) relies on centralized data aggregation,  which requires local devices to send their raw data to a central parameter server (PS). The conventional centralized approach is often impractical owing to communication bandwidth constraints and privacy concerns. Federated learning (FL) has thus been proposed as a distributed alternative, allowing edge devices to cooperatively train a global model while keeping local data private, which enhances privacy preservation and lowers communication costs~\cite{FL_beni,Min}.

In federated learning, the training procedure entails repeated exchanges of model parameters between the PS and local devices, which causes considerable communication overhead, particularly when a large number of local devices participate in the training process~\cite{Comm_res}.
In scenarios with limited communication capacity, such overhead often becomes a bottleneck that degrades training efficiency and constrains the scalability of FL systems~\cite{OTA_Sur}. 
To alleviate this issue, over-the-air (OTA) computation has been introduced as a promising technique for integrating communication and aggregation in FL~\cite{OTA1}. By exploiting the waveform superposition feature of wireless multiple-access channels, OTA computation aggregates local models from different devices concurrently, enabling low-latency and scalable global model aggregation \cite{OTAFL1, OTAFL2}. It has been verified that OTA computation can provide reduced latency and strong noise tolerance compared to conventional orthogonal multiple access methods, thus achieving scalable and prompt global model aggregation for FL \cite{OTA2}.

Due to the above significant advantages, OTA-FL has been widely investigated. Most of the existing designs are conducted under the ideal assumption of perfect channel state information (CSI)~ \cite{per3,per_CSI2,per_CSI3}. In practice, accurately estimating CSI remains a major challenge due to the time-varying nature of wireless channels and the large number of participating local devices \cite{time_vary}. Many existing studies have thoroughly examined how imperfect CSI affects the design and performance of traditional wireless communication systems \cite{Con_Per_CSI1, Con_Per_CSI2, Con_Imp_CSI, Con_Imp_CSI1}, however, the learning performance of OTA-FL under imperfect CSI conditions during local parameter uploading has been investigated by only a limited number of prior studies \cite{Imper_CSI, Related1, Related2, Related3, Related4}. For example, an OTA computation system under imperfect CSI is studied in \cite{Related2}, where the adverse impacts of channel estimation errors on OTA aggregation are analyzed.  To mitigate the impact of CSI imperfections, an adaptive retransmission strategy was proposed to reduce the MSE of the OTA-aggregated model. In \cite{Related3}, the
problem of computation distortion minimization is considered by taking an additive bounded uncertainty
of CSI into account, where the MSE of OTA aggregation is then minimized by jointly optimizing the transmitter and receiver design. Finally, the authors of \cite{Related4} use deep reinforcement learning method to solve the dynamic resource allocation problem for OTA-FL systems with imperfect CSI. 


Prior studies have primarily evaluated OTA-FL performance based on the instantaneous mean squared error (MSE) of the global model aggregated within a single round.
However, this approach neglects a key characteristic of OTA-FL, i.e., its iterative training nature, where aggregation errors accumulated across multiple rounds jointly affect the final learning outcome~\cite{Yinan,per_CSI1}. 
The work in~\cite{Yinan} established theoretical bounds on the temporal average of gradient norms, providing insights into the convergence behavior of OTA-FL architectures.
Similarly,~\cite{per_CSI1} investigated the gap between the loss of the final-round model and that of the optimal model, 
and proposed a joint transmitter–receiver optimization scheme to minimize this optimality gap. 
Nevertheless, these studies analyze the long-term performance of OTA-FL 
under idealized CSI assumptions, without accounting for practical channel estimation uncertainties.

In the context of OTA-FL with imperfect CSI, there exists a research gap in optimizing OTA-FL performance over multiple global model aggregation rounds. To bridge this research gap, we focus on the \textit{long-term MSE minimization} problem for global model aggregation under \textit{imperfect CSI} conditions. 
The main contributions of our work are summarized as follows.
\begin{itemize}
    \item We study an OTA-FL framework consisting of multiple single-antenna local devices and a multi-antenna PS. Unlike existing works that assume perfect CSI, we consider a more practical setting with imperfect channel knowledge. Owing to channel estimation errors, local devices upload their model parameters to the PS based on imperfect CSI. Correspondingly, both the transmit coefficient designs at the local devices and the receive combining vector optimizations at the central PS are carried out under imperfect CSI conditions.
    
    \item 
    Leveraging the proposed system architecture, we investigate the long-term accumulation of aggregation errors induced by imperfect CSI across multiple communication rounds, rather than focusing solely on per-round performance degradation. Specifically, we develop a unified analytical formulation that captures how channel estimation uncertainty propagates through OTA aggregation and gradually affects the global model updates. Under this framework, we establish a tractable MSE characterization that explicitly links imperfect CSI, aggregation distortion, and the resulting learning convergence behavior, thereby providing deeper insight into the performance limits of OTA-FL systems under realistic channel conditions.

    \item An optimization framework is proposed to minimize the long-term MSE of OTA model aggregation under imperfect channel conditions, through the joint design of transmit coefficients at the local devices and the receive combining vector at the PS. To enhance tractability, the original joint optimization problem is decomposed into two subproblems: (i) one targeting the design of transmit coefficients at the local devices, and (ii) the other focusing on the receive combining vector at the PS. We then reformulate each subproblem into a convex optimization problem and derive the closed-form solutions to reduce the design complexity.

 \item 
To enable online implementation without requiring future CSI, we develop a Lyapunov optimization-based causal transmit coefficient design that uses only the CSI available in the current round. A device-specific virtual queue
is introduced to track the cumulative deviation of each device's transmit power from its long-term average power budget. Based on the resulting drift-plus-penalty formulation, the transmit coefficients and receive combining vector are jointly optimized in each communication round. The proposed causal scheme enforces the long-term average power constraints while closely approaching the performance of the non-causal benchmark.

\item
Finally, extensive numerical experiments are conducted on the
Fashion-MNIST, CIFAR-10, and CIFAR-100 datasets under both IID and non-IID data distributions to validate the effectiveness and robustness of the proposed methods. We separately report the aggregation MSE and test accuracy to distinguish communication-layer
aggregation fidelity from end-to-end learning performance. We also compare our proposed schemes with other benchmarks, corresponding results demonstrate that the proposed schemes significantly mitigate the test accuracy degradation caused by imperfect CSI and outperform the considered benchmark schemes. 
\end{itemize}



\section{System Model} \label{sec:sysmodel}

\begin{figure}[t] 
\centering
\includegraphics[width=0.95\linewidth,trim=12 10 12 10,clip]{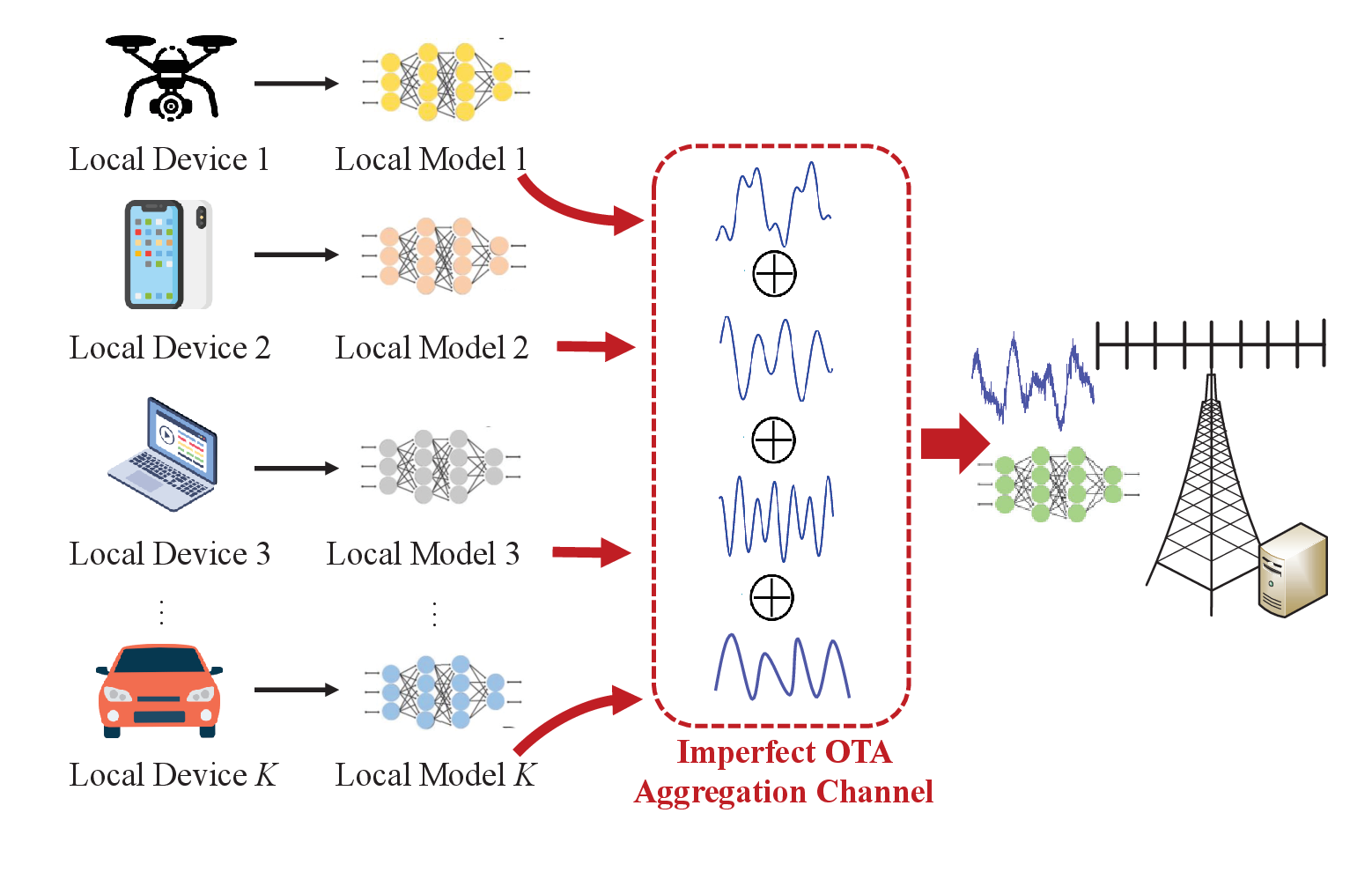}
\caption{Considered OTA-FL model with imperfect OTA aggregation channels, where multiple single antenna local devices upload parameters to the multi-antenna PS with imperfect CSI.}
\label{system_model}
\end{figure}

We study an OTA-FL system in which single-antenna local devices upload their model parameters to a multi-antenna PS to jointly train a global model, as shown in Fig. \ref{system_model}. We consider a practical scenario where the OTA aggregation channels from local devices to PS are imperfect, i.e., there exists channel estimation error, thereby the transmit coefficients and the receive combining vector can only be designed according to imperfect CSI. Under this practical scenario, we aim to minimize the long-term global model aggregation MSE through joint transmit coefficients and receive combining vector design to guarantee the final OTA-FL performance. 

\subsection{ Local Learning Model}
We study an OTA-FL system where the PS is co-located with a multi-antenna base station (BS).
A group of $K$ single-antenna edge devices collaboratively trains a common machine learning model.
The set of device indices is denoted by \(\mathcal{K} = \{1, 2, \ldots, K\}\). 
Each device \(k\) holds a local dataset \(\mathcal{D}_k\) with size \(|\mathcal{D}_k|\). 
At the \(k\)-th device, the local loss function, which depends on the model parameter vector \(\boldsymbol{w}_k \in \mathbb{R}^{N \times 1}\) is
\begin{equation}
F_k(\boldsymbol{w}_k; \mathcal{D}_k) \!\!= \!\!\frac{1}{|\mathcal{D}_k|} \sum_{i=1}^{|\mathcal{D}_k|} f(\boldsymbol{w}_k; (\boldsymbol{x}_{k,i}, y_{k,i})), \,\, for\,\, \forall k \in \mathcal{K},
\end{equation} 
where the loss function $f(\boldsymbol{w}_k; (\boldsymbol{x}_{k,i}, y_{k,i}))$ evaluates the model performance on training instance $i$ from node $k$, with $\boldsymbol{x}_{k,i}$ and $y_{k,i}$ representing the features and targets, respectively.
For analytical simplicity, it is assumed that all devices have the same number of local samples~\cite{Yinan}, 
and we let \( |\mathcal{D}_{\text{tot}}| = \sum_{k=1}^{K} |\mathcal{D}_k| \) represents the global dataset. 
Accordingly, the overall loss function can be written as
\begin{equation}
    F(\boldsymbol{w}; \mathcal{D}_{\text{tot}}) \!\!=\!\! \frac{1}{|\mathcal{D}_{\text{tot}}|} \!\!\sum_{k=1}^{K} \!|\mathcal{D}_k| F_k(\boldsymbol{w}_k; \mathcal{D}_k) 
\!\!= \!\!\frac{1}{K}\!\! \sum_{k=1}^{K} F_k(\boldsymbol{w}_k; \mathcal{D}_k).
\end{equation}
The learning objective is to obtain the optimal model parameter $\boldsymbol{w}^*$ that minimizes the global objective function, formally defined as
$\boldsymbol{w}^* = \arg\min_{\boldsymbol{w}} F(\boldsymbol{w}; \mathcal{D}_{\text{tot}})$.

\subsection{Over-the-Air Federated Learning}
To enable communication-efficient FL, we consider an OTA model aggregation scheme. 
In this approach, all edge devices perform several local training iterations 
to compute their individual gradients and then simultaneously upload 
their local model parameters to the PS via OTA transmission. 
The detailed computation and communication procedures are described as follows.

\textbf{1) Global Model Broadcast:} 
At the start of each global iteration \(t\), where \(t = 1, 2, \ldots, T\), 
the PS transmits the current global model \(\boldsymbol{w}(t) \in \mathbb{R}^{N \times 1}\) 
to all participating local devices. The downlink model broadcast from the PS to local devices is assumed to be reliable. This assumption is widely adopted in OTA-FL literature, since the PS is typically co-located with a base station and has significantly stronger transmission capability compared to edge devices. As a result, the broadcast of the aggregated model is often implemented via robust digital transmission, thereby it is reasonable to assume
that the distortion of the received global model at each local device is negligible \cite{negligible_distoration,Related1,Yinan1}.

\textbf{2) Local Gradient Update:} 
After obtaining the global parameter vector $\boldsymbol{w}(t)$, 
every participating device $k \in \{1, 2, \ldots, K\}$ 
sets its initial parameters to $\boldsymbol{w}_k(t,0) = \boldsymbol{w}(t)$. 
Subsequently, each device performs \(I \geq 1\) local update iterations 
based on its own stochastic gradient, which can be expressed as
\begin{equation} \label{local_model}
    \boldsymbol{w}_k(t, i + 1) = \boldsymbol{w}_k(t, i) - \lambda \tilde{\boldsymbol{g}}_k(t, i), \quad i = 0, \ldots, I - 1,
\end{equation}
where \(\lambda\) denotes the learning rate, 
\(\boldsymbol{w}_k(t, i + 1)\) represents the local model of device \(k\) 
in the \(t\)-th global round after performing \(i\) local update steps, and
\begin{align}
     \tilde{\boldsymbol{g}}_k(t, i) &= \nabla F_k\left(\boldsymbol{w}_k(t, i); \mathcal{B}_k(t, i)\right) \nonumber\\
&= \frac{1}{B} \sum_{(\boldsymbol{x}_k, y_k) \in \mathcal{B}_k(t, i)} \nabla f(\boldsymbol{w}_k(t, i); (\boldsymbol{x}_k, y_k))
\end{align}
is the stochastic gradient computed based on the mini-batch 
\(\mathcal{B}_k(t, i)\), which consists of \(B\) data samples 
randomly drawn from the local dataset \(\mathcal{D}_k\).

\textbf{3) OTA Global Model Aggregation:}  
To construct the global model, each device uploads its locally updated parameters to the PS.  
In the uploading stage, every device concurrently sends its computed local gradient updates to the PS following appropriate signal preparation. Exploiting the additive nature of multi-access channels, the PS obtains a combined signal that represents the sum of all local gradient updates. Therefore, the upload latency stays constant as devices scale, facilitating efficient model updates across large FL networks.

For the purpose of controlling transmit power at each local device, the $N$-dimensional local model update vector is first
defined as
\begin{equation}
    \boldsymbol{\theta}_k (t)=\frac{\boldsymbol{w}_k(t,0)-\boldsymbol{w}_k(t,I)}{\lambda}=\sum_{i=0}^{I-1}\tilde{\boldsymbol{g}}_k(t, i).
\end{equation}
After obtaining the local model, each device \(k\) computes the mean value \(\bar{\theta}_k(t)\) 
and the variance \(\pi_k^2(t)\) of the local gradient vector \(\boldsymbol{\theta}_k(t)\) as
\begingroup
\setlength{\jot}{0pt}
\vspace{-2mm}
\begin{align}
\bar{\theta}_k(t) 
&= \frac{1}{N}\sum_{n=1}^{N}\theta_{k,n}(t),
\quad \forall k \in \mathcal{K}, \label{eq:theta_mean}\\[-0.5mm]
\pi_k^2(t) 
&= \frac{1}{N}\sum_{n=1}^{N}
\left(\theta_{k,n}(t)-\bar{\theta}_k(t)\right)^2,
\quad \forall k \in \mathcal{K}. \label{eq:theta_var}
\end{align}
\endgroup
here, \(\theta_{k,n}(t)\) denotes the \(n\)-th entry of the vector \(\boldsymbol{\theta}_k(t)\). The global empirical mean and variance across all participating
devices and model coordinates are then obtained from these local
statistics as
$
\bar{\theta}(t)
=
\frac{1}{K}\sum_{k\in\mathcal{K}}\bar{\theta}_k(t)$,
and
$
\pi^2(t)
=
\frac{1}{K}\sum_{k\in\mathcal{K}}
\left[
\pi_k^2(t)
+
\left(\bar{\theta}_k(t)-\bar{\theta}(t)\right)^2
\right]$,
respectively. Each device then applies the same affine
normalization according to
\begin{equation}
s_{k,n}(t)
=
\frac{\theta_{k,n}(t)-\bar{\theta}(t)}{\pi(t)},
\quad
n=1,2,\ldots,N,\quad \forall k\in\mathcal{K}.
\end{equation}
Under this global normalization, an individual normalized update
vector $\mathbf{s}_k(t)$ is not required to have exactly zero
empirical mean and unit empirical variance. Instead, the collection
of normalized symbols across all devices and model coordinates
satisfies
$
\frac{1}{KN}
\sum_{k\in\mathcal{K}}\sum_{n=1}^{N}s_{k,n}(t)=0,
$
and
$
\frac{1}{KN}
\sum_{k\in\mathcal{K}}\sum_{n=1}^{N}
\left|s_{k,n}(t)\right|^2=1
$.
Unlike the device-wise normalization in~\cite{R22},
which relies on device-specific statistics, we employ a common
global mean and variance across all participating devices to preserve a uniform OTA summation structure and enable common post-aggregation de-normalization at the PS side. This normalization relies on empirical statistics estimated from high-dimensional
model updates, the normalized representation becomes approximately zero-mean with approximately unit
variance due to the self-averaging effect when the model dimension $N$ is large \cite{per_CSI1,normalize0,normalize1}.
The normalization of gradients can be realized by exchanging their statistical information, such as the mean $\{\bar{\theta}_k(t)\}$ and the standard deviation $\{\pi_k(t)\}$, 
between the edge devices and the PS. 
The exchanged statistical quantities are assumed to be transmitted without error 
and incur negligible signaling overhead, 
since their data size is much smaller than that of the complete gradient vectors 
\cite{Yinan, Yinan1, Exchange1, Exchange2}.

We now describe the underlying uploading process for the global model update, where a $M$-antenna PS serves $K$ single-antenna local devices. We assume that the channel coefficients remain the same within each communication round $t$, and we consider the imperfect channel estimates. Specifically, the perfect channel from the local device $k$ to the PS is $\boldsymbol{h}_k(t) \in \mathbb{C}^{M \times 1}$, due to the imperfect channel estimations, the estimated channel $\hat{\boldsymbol{h}}_k(t)$ can be expressed as $\hat{\boldsymbol{h}}_k(t)=\boldsymbol{h}_k(t)+\tilde{\boldsymbol{h}}_k(t)$, where $\tilde{\boldsymbol{h}}_k(t)$ represents the channel estimation error, and each entry of it has zero mean and variance $\sigma_h^2$ \cite{Chan_err1}. Without loss of generality, we assume that we use the minimum mean-squared-error (MMSE) method to the obtain the estimated CSI \cite{MRC_Com,inden}.
To counteract the impact of channel fading, $\boldsymbol{s}_k(t)$ is multiplied by a pre-processing coefficient $\mu_k(t)$ prior to transmission. Accordingly, the received signal at the PS can be formulated as
\begin{equation} \label{received_signal}
    \mathbf{Y}(t)=\sum_{k=1}^K\boldsymbol{h}_{k}(t)\mu_k(t)\boldsymbol{s}_{k}^T(t)+\mathbf{Z}(t),
\end{equation}
where $\mathbf{Z}(t) \in \mathbb{C}^{M \times N}$ represents the noise, each entry of it follows a circularly symmetric complex Gaussian distribution with zero mean and variance $\sigma_0^2$.
With $M$ receiving antennas at the PS, we are able to apply a combining vector $\boldsymbol{b}(t) \in \mathbb{C}^{M \times 1}$ to properly combine the received local parameters for better global model aggregation, where the estimated global model can be expressed as
\setlength{\jot}{1pt}
\begin{align}\label{received-signal0}
  \hat{\boldsymbol{s}}^T
  (t)&=\boldsymbol{b}^H(t)\mathbf{Y}
  (t)\nonumber\\  &=\!\!\!\sum_{k=1}^K\boldsymbol{b}^H(t)\left(\hat{\boldsymbol{h}}_{k}(t)-
\tilde{\boldsymbol{h}}_k(t)\right)    \mu_k(t)\boldsymbol{s}^T_{k}(t)+\boldsymbol{b}^H(t)\mathbf{Z}(t).  
\end{align} 
The instaneous MSE is thereby defined as
\setlength{\jot}{1pt}
\begin{align} \label{ins-MSE}
   &\text{MSE}(t)=\mathbb{E}\lbrace||
\boldsymbol{e}^T(t)
||^2\rbrace=\mathbb{E}\lbrace||
\hat{\boldsymbol{s}}^T(t) -\boldsymbol{s}^T(t)
||^2 \rbrace\nonumber\\
&= N
\bigg( \sum_{k=1}^{K} 
\left| \boldsymbol{b}^H(t) \hat{\mathbf{h}}_k(t) \mu_k(t) - 1 \right|^2 
+\sum_{k=1}^{K} |\mu_k(t)|^2 \sigma_h^2 \left\| \boldsymbol{b}(t) \right\|^2 \nonumber\\
& \quad\quad\quad\quad\quad\quad\quad\quad\quad\quad\quad\quad\quad\quad\quad+ \left\| \boldsymbol{b}(t) \right\|^2 \sigma_0^2 
\bigg).
\end{align}
Here, $\boldsymbol{s}^T(t)=\sum_{k=1}^K\boldsymbol{s}^T_k(t)$ represents the perfect global model aggregation results.
The MSE decomposition in \eqref{ins-MSE} is obtained by substituting \eqref{received-signal0} into the error definition and expanding the expectation. Following standard OTA-FL assumptions, the normalized local updates are modeled as uncorrelated and independent of receiver noise~\cite{normalize0,normalize1,Assumption}, so the cross terms vanish and the MSE separates into signal misalignment, imperfect-CSI-induced distortion, and receiver noise.
Specifically, in \eqref{ins-MSE}, the first term $\sum_{k=1}^{K} 
\left| \boldsymbol{b}^H(t) \hat{\mathbf{h}}_k(t) \mu_k(t) \!-\! 1 \right|^2$ represents the signal misalignment error, the second term  $\sum_{k=1}^{K} |\mu_k(t)|^2\sigma_h^2 \left\| \boldsymbol{b}(t)\right\|^2$ denotes the CSI-related error, and the third term $\left\| \boldsymbol{b}(t) \right\|^2 \sigma_0^2$ is the noise introduced error. As shown in \eqref{ins-MSE}, the transmit coefficients $\mu_k(t),\, \forall\, k \in \mathcal{K}, \, t=1,2,...,T$ and the combining vector $\boldsymbol{b}(t), \, t=1,2,...,T$, substantially affect the MSE of global model aggregation. Hence, it is essential to design them judiciously to minimize the global model aggregation error.


The OTA-FL system repeatedly executes the steps outlined in \eqref{local_model} – \eqref{received-signal0} until reaching model convergence or exhausting the specified maximum iteration.
To ensure energy efficiency throughout the OTA-FL training process,  
each device is subject to both maximum and average transmit power constraints, 
which can be expressed as
\begin{equation}
    |\mu_k(t)|^2 \leq P_k^{max}, \,\,\forall \, k=1,2,...,K, \,\, \,\,t=1,2,...,T,
\end{equation}
and
\begin{equation}
    \frac{1}{T}\sum_{t=0}^{T-1} |\mu_k(t)|^2 \leq P_k^{ave}, \,\,\forall \, k=1,2,...,K.
\end{equation}
In this setup, $P_k^{\max} \ge 0$ and $P_k^{\text{ave}} \ge 0$ represent the maximum and average transmission power limits for the $k$-th device, while $T$ denotes the overall iteration count for model updates.
For consistency, it is assumed that the average transmit power does not exceed the maximum transmit power, i.e., \(P_k^{\text{ave}} \le P_k^{\max}\), for every device \(k\)~\cite{Yinan, Yinan1}.

\vspace{-0.2cm}
\section{Problem Formulation for Long-term MSE Minimization in OTA-FL Systems} \label{sec:convanalysis}
In this work, we focus on the long-term performance analysis and design of OTA-FL systems.  
Accordingly, this section first presents a convergence analysis that accounts for 
multiple global aggregation rounds and the impact of channel estimation errors.  
Next, we develop an optimization framework to minimize the theoretical upper limit on the time-averaged MSE for the global model.
\vspace{-0.2cm}
\subsection{Preliminary}
\subsubsection{Non-Independent and Identically Distributed (non-IID) Data} 
When the datasets across local devices are non-IID, individual loss functions can have optima that differ from the overall objective's minimum \cite{Xiaoyan,Fangtong}. As gradient divergence indicates data heterogeneity, we have Definition 1.

\textit{Definition 1:} 
Let $\{\nabla F_k(\boldsymbol{w})\}_{k=1}^{K}$ denote the local gradients of $K$ edge devices. 
We use $\sigma_d^2$ to characterize the degree of gradient heterogeneity, defined as
\begin{equation}
    \left\| \nabla F_k(\boldsymbol{w}) - \nabla F(\boldsymbol{w}) \right\|_2^2 \leq \sigma_d^2, 
    \quad \forall k \in \mathcal{K}.
\end{equation}
Note that the distance between any local gradient $\nabla F_k(\boldsymbol{w})$ and the global gradient 
$\nabla F(\boldsymbol{w})$ is constrained by $\sigma_d^2$.
Under the IID setting, local gradients become identical as the number of samples grows large, 
leading to \(\sigma_d^2 = 0\). 
In contrast, when data distributions differ among devices, the corresponding local gradients exhibit directional divergence, 
reflecting statistical heterogeneity across the network.

\textit{Assumption 1 (Lower-Bounded Loss Function):} The loss function $F(\boldsymbol{w})$ is bounded below, i.e., $F(\boldsymbol{w}) \geq F(\boldsymbol{w}^*) > -\infty$, where $\boldsymbol{w}^*$ 
denotes the global minimizer.


\textit{Assumption 2 (Lipschitz Continuity and Smoothness):}  
Each local loss function \(F_k(\boldsymbol{w})\) is assumed to be continuously differentiable 
and \(L\)-smooth, where \(L > 0\) is a non-negative constant such that  
\begin{equation}
    \left\| \nabla F_k(\boldsymbol{w}) - \nabla F_k(\boldsymbol{w}') \right\|_2 
    \leq L \left\| \boldsymbol{w} - \boldsymbol{w}' \right\|_2, 
    \quad \forall\, \boldsymbol{w}, \boldsymbol{w}'.
\end{equation}
This smoothness condition implies the following inequality:
\begin{equation}
    F_k(\boldsymbol{w}') 
    \leq F_k(\boldsymbol{w}) 
    + \left\langle \nabla F_k(\boldsymbol{w}), \boldsymbol{w}' - \boldsymbol{w} \right\rangle 
    + \frac{L}{2} \left\| \boldsymbol{w} - \boldsymbol{w}' \right\|_2^2.
\end{equation}


\textit{Assumption 3 (Bounded Variance of Stochastic Gradients):}  
The stochastic mini-batch gradients $\{ \tilde{\boldsymbol{g}}_k \}$ 
are considered independent and unbiased approximations of the true gradients 
$\{ \nabla F_k(\boldsymbol{w}_k) \}$, and their variance is assumed to be bounded as
\begin{align}
   & \mathbb{E}[\tilde{\boldsymbol{g}}_k] = \nabla F_k(\boldsymbol{w}_k), 
   \quad \forall k \in \mathcal{K}, \\
   & \mathrm{Var}(\tilde{\boldsymbol{g}}_k) 
   = \mathbb{E} \left[ \left\| \tilde{\boldsymbol{g}}_k 
   - \nabla F_k(\boldsymbol{w}_k) \right\|_2^2 \right] 
   \leq \xi^2, \quad \forall k \in \mathcal{K},
\end{align}
where \(\xi \geq 0\) is a constant representing the upper bound on the stochastic gradient variance, 
which captures the sampling noise introduced by mini-batch training.

\textit{Assumption 4 (Limited Variance):}  
For every element within $\boldsymbol{\theta}_k$, its variance is assumed to remain below a non-negative bound $\Gamma$,  
that is, $\pi_k^2 \leq \Gamma$.

Note that Assumptions~1 - 4 and Definition~1 are standard regularity conditions widely used in federated learning convergence analysis \cite{OTAFL1,normalize0, Assumption}. Definition~1 characterizes bounded statistical heterogeneity across local devices. Assumption~1 holds for common non-negative learning objectives, such as cross-entropy and mean-squared-error losses, while Assumption~2 is the standard Lipschitz-smoothness condition. Assumption~3 follows from unbiased mini-batch stochastic gradient estimation with independent sampling and finite variance. Assumption~4 is ensured by the normalization step and transmit-power constraints, which bound the energy and variance of transmitted local updates.

\subsection{Convergence Analysis for the Considered OTA-FL System}
Under the preceding assumptions, we formulate the following theorem to characterize the convergence performance of the OTA-FL framework in the presence of non-IID data and multiple local update steps.

\textit{\textbf{Theorem 1}}: With Definition 1 and Assumptions 1 - 4, if
learning rate $\lambda$ satisfies $\lambda \leq \min \left\{ \frac{1}{2LI},\ \frac{1}{\sqrt{6 L^2I^3}},\ \frac{\sqrt{I - 1}}{4LI} \right\}$, then the temporal mean of gradient norms following $T$ iterations of model aggregation is bounded by
\begin{align} \label{OBJ0}
&\frac{1}{T} \sum_{t=0}^{T-1} \left\| \nabla F(\mathbf{w}(t)) \right\|_2^2
\leq 
\underbrace{\frac{4 \left(F(\mathbf{w}(0)) - F(\mathbf{w}^*)\right)}{\lambda(I - 1) T}}_{\text{Initial optimality gap}} \nonumber \\
&\quad + \underbrace{\frac{4\left(4\lambda^2 L^2 I^3 + 3I\right)}{3(I - 1)} \xi^2}_{\text{Variance of stochastic gradient}} 
+ \underbrace{\frac{16 \lambda^2 L^2 I^3}{I - 1} \sigma_d^2}_{\text{Gradient divergence}} \nonumber \\
&\quad + \underbrace{\frac{2(1 + 2\lambda L)}{I - 1}  \frac{\Gamma}{K^2}  \frac{1}{T} \sum_{t=0}^{T-1} \text{MSE}(t)}_{\text{Time-averaged MSE}},
\end{align}
where $\text{MSE}(t)$ is the instaneous MSE for global model aggregation round $t$ as we previously defined in \eqref{ins-MSE}.

\vspace{0.05 in}
\textit{Proof}: See Appendix \ref{Theorem1}. \qed
\vspace{0.05 in}

From the above formulation, we can see that the channel estimation error directly enters the received signal model in \eqref{received-signal0}, giving rise to the CSI-induced term in the MSE expression \eqref{ins-MSE}. This imperfect-CSI-aware MSE is then incorporated into the long-term convergence bound in \eqref{OBJ0}.
Specifically, the first three terms in \eqref{OBJ0} correspond to the initial optimality gap, stochastic gradient noise, and gradient heterogeneity, respectively, while the last term captures the time-averaged MSE caused by non-ideal aggregation channels. As \(T \rightarrow \infty\), the initial optimality gap vanishes, and for fixed local devices, local steps, and step size, the second and third terms remain unchanged. Therefore, improving OTA-FL convergence mainly depends on minimizing the time-averaged MSE term in \eqref{OBJ0}.

\vspace{-1 em}
\subsection{Problem Formulation}
Omitting constants in  \eqref{OBJ0}, the time-averaged MSE becomes
\begin{align} \label{bar_MMSE}
   & \overline{\text{MSE}}= \sum_{t=0}^{T-1}
\bigg( \sum_{k=1}^{K} 
\left| \boldsymbol{b}^H(t) \hat{\mathbf{h}}_k(t) \mu_k(t) - 1 \right|^2 \nonumber\\& \quad\quad\quad\quad\quad+\sum_{k=1}^{K} |\mu_k(t)|^2 \sigma_h^2 \left\| \boldsymbol{b}(t) \right\|^2 + \left\| \boldsymbol{b}(t) \right\|^2 \sigma_0^2 
\bigg).
\end{align}

The time-averaged MSE is minimized through the joint optimization of the transmit coefficient $\mu_k(t)$ at each local device and the receive combining vector $\boldsymbol{b}(t)$ at the PS. The resulting optimization problem is formulated as:
\begin{align}
 (\mathcal{P}0): \quad 
&\min_{\mu_k(t),\boldsymbol{b}(t) } \ \overline{\mathrm{MSE}} \\
&\quad\quad\text{s.t.}\quad
 |\mu_k(t)|^2 \leq P_k^{max}, \,\, \quad \forall k, \forall t,  \label{Max_Power_Cons}\\
&\quad\quad\quad\quad\frac{1}{T}\sum_{t=0}^{T-1} |\mu_k(t)|^2 \leq P^{ave}_k,  \quad \forall k.  \label{aver_Power_Cons}
\end{align}

Our design goal is to minimize the time-averaged MSE by optimizing the transmit coefficient $\mu_k(t)$ at the local device and the combining vector $\boldsymbol{b}(t)$ at the PS. The constraint \eqref{Max_Power_Cons} is the maximum power constraint for each local device in each global model upload round, and \eqref{aver_Power_Cons} is the long-term average power constraint for each local device. Note that $\boldsymbol{b}(t)$ is the receive combining vector, hence its norm is not restricted \cite{Constraint}.

\section{Alternative Optimization for Minimizing Time-averaged MSE} \label{sec:alternative}
We now address~$(\mathcal{P}0)$ by alternately optimizing its interdependent variables and provide analytic solutions for each subproblems. Specifically, we first analyze the transmit coefficients design problem at the local device side, then we derive the combining vector design at the PS side.

\subsection{Transmit Coefficients Design at the Local Devices}
First, we fix the combining vector design and optimize the transmit coefficient $\mu_{k}(t)$ by solving the following problem
\begin{align}
(\mathcal{P}1) : & \min_{\mu_{k}(t),\,k=1,2,..,K} \quad \overline{\mathrm{MSE}} \\
    &\quad\text{s.t.}\quad
 |\mu_k(t)|^2 \leq P_k^{max}, \,\, \quad \forall k, \forall t,   \\
&\quad\quad\quad\frac{1}{T}\sum_{t=0}^{T-1} |\mu_k(t)|^2 \leq P^{ave}_k,  \quad \forall k. 
\end{align}

 With $K$ local devices in the network, we partition $(\mathcal{P}1)$ into $K$ independent subproblems to determine the transmission parameters for each device through
\begin{align}
(\mathcal{P}1.1) : & \min_{\mu_{k}(t)}  \quad\quad \overline{\mathrm{MSE}}_k \left(\mu_{k}(t)\right)\\
    &\quad\text{s.t.}\quad\,\,\,\,
|\mu_k(t)|^2 \leq P_k^{max}, \,\, \quad \forall t,\label{indi_power} \\
&\quad\quad\quad\,\,\,\,
\frac{1}{T}\sum_{t=0}^{T-1} |\mu_k(t)|^2 \leq P^{ave}_k.  \label{aver_power}
\end{align}
where 
\begin{align}
&\overline{\mathrm{MSE}}_k\left(\mu_{k}(t)\right)=\\
&\sum_{t=0}^{T-1}
\bigg( 
\left| \boldsymbol{b}^H(t) \hat{\mathbf{h}}_k(t) \mu_k(t) - 1 \right|^2 +|\mu_k(t)|^2 \sigma_h^2 \left\| \boldsymbol{b}(t) \right\|^2 \nonumber
\bigg). 
\end{align}
For a given receive combining vector $\boldsymbol{b}(t)$, we first
optimize the phase of the transmit coefficient $\mu_k(t)$. Let $
a_k(t) =\boldsymbol{b}^H(t)\hat{\boldsymbol{h}}_k(t)$,
and write $\mu_k(t)=r_k(t)e^{j\phi_k(t)}$, where $r_k(t)\geq 0$.
The per-round MSE term corresponding to device $k$ can then be expressed as
\begin{align}
&\left|a_k(t)\mu_k(t)-1\right|^2
+|\mu_k(t)|^2\sigma_h^2\|\mathbf{b}(t)\|^2=
\nonumber\\
& \quad \quad
\left(|a_k(t)|^2+\sigma_h^2\|\mathbf{b}(t)\|^2\right)r_k^2(t) \nonumber\\
&\quad \quad\quad \quad  -2|a_k(t)|r_k(t)
\cos\!\left(\angle a_k(t)+\phi_k(t)\right)+1.
\end{align}
Thus, for any fixed amplitude $r_k(t)$, the MSE is minimized
when $\phi_k(t)=-\angle a_k(t)$. Therefore, without loss of
optimality, the transmit coefficient can be parameterized as
$\mu_k(t)=\sqrt{P_k(t)}
\frac{\hat{\mathbf{h}}_k^H(t)\mathbf{b}(t)}
{\left|\mathbf{b}^H(t)\hat{\mathbf{h}}_k(t)\right|}$,
where $P_k(t)=|\mu_k(t)|^2$ denotes the transmit power of
device $k$ in round $t$. This phase-alignment structure is
consistent with the transmit pre-processing adopted in
\cite{Trans_Coe1,Yinan1,R11}. Consequently, $(\mathcal{P}1.1)$ can be
equivalently reformulated as
\begin{align}
(\mathcal{P}1.2) : & \min_{P_{k}(t)}  \quad\quad \overline{\mathrm{MSE}}_k \left(P_{k}(t)\right)\\
    &\quad\text{s.t.}\quad\,\,\,\,
0 \leq P_k(t) \leq P_k^{max}, \,\, \label{indi_power} \\
&\quad\quad\quad\,\,\,\,
\frac{1}{T}\sum_{t=0}^{T-1} P_k(t) \leq P^{ave}_k.  \label{aver_power}
\end{align}
where 
\setlength{\jot}{1pt}
\begin{align}
&\overline{\mathrm{MSE}}_k\left(P_{k}(t)\right)=\\
&\sum_{t=0}^{T-1}
\bigg( 
\left( | \mathbf{b}^H(t) \hat{\mathbf{h}}_k(t)|\sqrt{P_k(t)} - 1 \right)^2 +P_k(t) \sigma_h^2 \left\| \mathbf{b}(t) \right\|^2 \nonumber
\bigg). 
\end{align}
After analytically determining the optimal phase of the transmit coefficient, it remains to optimize its amplitude, i.e., the
transmit power $P_k(t)$.
And we have the following theorem regarding to the optimal transmit power $P_k^*(t)$ of device $k$ in round $t$.


\textbf{\textit{Theorem 2}}: The optimal transmit power $P_k^*(t)$ of user $k$ in global model aggregation round $t$ is calculated based on the following criterion:

\begin{itemize}
    \item If condition
    \begin{equation} \label{power_condi}
      \sum_{t=0}^{T-1}\!\!\min\!\! \left\lbrace\!\!\!
    \left( \frac{\left| \boldsymbol{b}^H(t) \hat{\boldsymbol{h}}_k(t) \right|}{\left| \boldsymbol{b}^H(t) \hat{\boldsymbol{h}}_k(t) \right|^2 \!\!+\! \sigma_h^2 \|\boldsymbol{b}(t)\|^2} \!\!\right)^2\!\!\!\!, P_k^{max}\!\!
    \right\rbrace \!\! \leq\! T P^{ave}_k  
    \end{equation}
    holds, the analytical solution for the optimal transmission power $P_k^*(t)$ in problem~$(\mathcal{P}1.2)$ is given by
    \begin{equation} \label{power1}
      P_k^*(t)\!=\!  \min \!\!\left\lbrace\!\!\!
    \left(\!\! \frac{\left| \boldsymbol{b}^H(t) \hat{\boldsymbol{h}}_k(t) \right|}{\left| \boldsymbol{b}^H(t) \hat{\boldsymbol{h}}_k(t) \right|^2 + \sigma_h^2 \|\boldsymbol{b}(t)\|^2} \!\!\right)^2\!\!\!\!, P_k^{\text{max}}\!\!
    \right\rbrace.
    \end{equation}
    \item Otherwise, the optimal transmit power is given by
    \begin{equation}\label{power2}
      P_k^*(t)\!\!=\!\!   \min \!\! \left\lbrace\!\!\!
    \left( \!\!\frac{\left| \boldsymbol{b}^H(t) \hat{\boldsymbol{h}}_k(t) \right|}{\left| \boldsymbol{b}^H(t) \hat{\boldsymbol{h}}_k(t) \right|^2\!\! + \!\sigma_h^2 \|\boldsymbol{b}(t)\|^2 \!\! + \!\!\rho_k^*} \!\!\right)^2\!\!\!\!, P_k^{\text{max}}
    \!\!\right\rbrace,
    \end{equation}
   where $\rho_k^*$ is obtained through a one-dimensional bisection procedure, 
ensuring that the average transmit-power constraint is satisfied.
\end{itemize}

\vspace{0.05 in}
\textit{Proof}: See Appendix \ref{Theorem2}.\qed

Accordingly, the optimal transmit coefficient is given by
$\mu_k^*(t)=
\sqrt{P_k^*(t)}
\frac{\hat{\boldsymbol{h}}_k^H(t)\boldsymbol{b}(t)}
{\left|\boldsymbol{b}^H(t)\hat{\boldsymbol{h}}_k(t)\right|}$.
From Theorem 2, we can observe that imperfect CSI affects $P_k^*(t)$ in two ways. First, the estimated channel $\hat{\boldsymbol{h}}_k(t)$ determines the estimated effective channel gain $\left|\boldsymbol{b}^H(t)\hat{\boldsymbol{h}}_k(t)\right|$, which is device-dependent. Second, the channel estimation error variance $\sigma_h^2$ appears as an uncertainty penalty in the denominator. A larger $\sigma_h^2$ increases the CSI-error-induced distortion term and thus changes the optimal power allocation.

\subsection{Combining Vector Design at the PS}
With updated transmit coefficients design at the local device, we then optimize the combining vector $\boldsymbol{b}(t) \in \mathbb{C}^{M \times 1}$ at the PS side as
\begin{align}
(\mathcal{P}2) : \min_{\boldsymbol{b}(t)}\quad \overline{\mathrm{MSE}},
\end{align}
where $\overline{\mathrm{MSE}}$ is defined in \eqref{bar_MMSE}. To solve this problem, we first decouple $(\mathcal{P}2)$ into $T$ subproblems for $T$ global model aggregation rounds, i.e.,
\begin{align}
     (\mathcal{P}2.1) : \min_{\boldsymbol{b}(t)}\quad \overline{\mathrm{MSE}}_t\left(\boldsymbol{b}(t)\right),
\end{align}
where 
\setlength{\jot}{1pt}
\begin{align}
&\overline{\mathrm{MSE}}_t\left(\boldsymbol{b}(t)\right)=
\sum_{k=1}^{K} 
\left| \mathbf{b}^H(t) \hat{\mathbf{h}}_k(t) \mu_k(t) - 1 \right|^2 \nonumber\\& \quad\quad\quad\quad\quad+\sum_{k=1}^{K} |\mu_k(t)|^2 \sigma_h^2 \left\| \mathbf{b}(t) \right\|^2 + \left\| \mathbf{b}(t) \right\|^2 \sigma_0^2 .
\end{align}
Furthermore, we have the following theorem regarding to the solution of $(\mathcal{P}2.1)$.

\textbf{\textit{Theorem 3}}: The optimal value $\boldsymbol{b}^*(t)$ of problem $(\mathcal{P}2.1)$ is:
\begin{equation} \label{b(t)}
   \boldsymbol{b}^*(t) =\left(
\sum_{k=1}^K\boldsymbol{\delta}_k (t)\boldsymbol{\delta}_k^H (t)+\beta\mathbf{I}_M
\right)^{-1}\sum_{k=1}^K\boldsymbol{\delta}_k(t),
\end{equation}
where $\boldsymbol{\delta}_k(t)=\hat{\boldsymbol{h}}_k (t)\mu_k(t)\in \mathbb{C}^{M \times 1}$, $\beta=\sum_{k=1}^K|\mu_k(t)|^2\sigma_h^2+\sigma_0^2$ and $\mathbf{I}_{M}$ represents the identity matrix of dimension $M \times M$.

\vspace{0.05 in}
\textit{Proof}: See Appendix \ref{Theorem3}. \qed
\vspace{0.05 in}

\begin{algorithm}[t]
\centering
\begin{minipage}{0.45\textwidth}  
\caption{Non-causal Imperfect CSI aided Long-Term MSE Minimization for OTA-FL}
\label{Alg1}

\textbf{Input:} Channel information for all T rounds $\{\hat{\boldsymbol{h}}_k(t)\}_{t=0}^{T-1}$; stopping condition $\epsilon$; maximum iteration time $Q$\;

\textbf{Initialization:} The transmit power at each local device, receive combining vector, and iteration index $q = 0$\;

\Repeat{$\dfrac{\overline{\mathrm{MSE}}_{q-1} - \overline{\mathrm{MSE}}_{q}}{\overline{\mathrm{MSE}}_q} \leq \epsilon$ or $q=Q$}{
    (i): $q=q+1$;\\
    (ii): Given the combining vector $\{\boldsymbol{b}(t)\}_{q-1}$, update $\{P_k(t)\}_{q}$ according to \eqref{power1} if \eqref{power_condi} holds; otherwise, update $\{P_k(t)\}_q$ via \eqref{power2}; \\
    (iii): Given $\{P_k(t)\}_{q}$, update $\{\boldsymbol{b}(t)\}_q$ via \eqref{b(t)};
}

\textbf{Output:} $\{P_k^*(t)\}$ for $k=1,2,...,K$, $t=1,2,...,T$ and $\{\boldsymbol{b}^*(t)\}$ for $t=1,2,...,T$.
\end{minipage}
\end{algorithm}

The original problem~$(\mathcal{P}0)$ can now be decomposed into sub-problems~$(\mathcal{P}1.2)$ and~$(\mathcal{P}2.1)$, which are solved in an alternating manner. The complete procedure of our proposed approach is detailed in Algorithm~1.
\textit{It is worth noting that, while Algorithm~1 is capable of solving the problem, it suffers from a key drawback: the design of transmit coefficients relies on non-causal CSI, i.e., the channel information of all OTA global aggregation rounds, to evaluate the criterion in~\eqref{power_condi}, which limits its practicality in real-world implementations.} To tackle this challenge, we further propose a Lyapunov optimization-based approach, which effectively decouples the long-term power constraint into individual constraints at each global model aggregation round.

\subsection{Lyapunov Optimization-based Transmit Coefficients Design at the PS Side}
In order to decouple the time-based coupling, we introduce the concept of virtual queues. Specifically, a virtual queue is established for each local device to represent the current cumulative energy consumption, which can be expressed as 
\begin{equation} \label{queue}
    q_k[t]=\text{max}\lbrace
    q_k[t-1] + P_k(t)-P^{ave}_k , 0
    \rbrace, \,\, q_k[0]=c_k,
\end{equation}
with $c_k \geq 0$. Specifically, $c_k>0$ means that the average power constraint has already been violated at the beginning of the learning process (i.e., an initial queue backlog exists), $c_k=0$ represents a neutral state with no accumulated violation. Note that the initial queue value can be any finite constant without affecting
the stability of the virtual queue or the validity of the Lyapunov
optimization framework  \cite{Lya,Lya1}.

According to \eqref{queue}, we have 
\begin{equation}
\frac{1}{T} \sum_{t=1}^{T} \left( P_k(t) - P_k^{ave}\right) 
\leq \frac{1}{T} \left( q_k[T] - q_k[0] \right),
\end{equation}
where the long-term power constraint can be met when $q_k[T]/T=0$. Based on the virtual queue, the Lyapunov drift function can be defined as
\begin{equation}
 \Delta L_k(t) = \mathbb{E} \left\{ \frac{1}{2} \left( q_k[t]^2 - q_k[t-1]^2 \right) \,\middle|\, q_k[t-1] \right\},   
\end{equation}
the Lyapunov drift plus penalty function can be defined as $\Delta L_k(t)+V\overline{\mathrm{MSE}}_k(P_k(t)) $, where the penalty factor $V >0$ is introduced to balance the average power stability and the global model aggregation efficiency of OTA-FL. A larger value of $V$ places more emphasis on minimizing the model aggregation error and generally results in a lower aggregation MSE. Conversely, a smaller value of $V$ enforces stricter average power constraints over iterations, thus, the power allocation adapts more quickly to meet the average power limit~\cite{Lya}. Based on~\eqref{queue}, we obtain
\begin{equation}
    \Delta L_k(t) \leq \frac{1}{2} \left( P_k(t) - P_k^{ave} \right)^2 
+ q_k[t-1] \left(  P_k(t) - P_k^{ave} \right).
\end{equation}
Therefore, we can reformulate a new optimization problem regarding the transmit power design:
\begin{align}
(\mathcal{P}1.3) : & \min_{P_{k}(t)}  \quad\quad \digamma(t)+V\mathrm{MSE}_k(P_k(t)), \label{obj1.3}\\
    &\quad\text{s.t.}\quad\,\,\,\,
0\leq P_k(t) \leq P_k^{max}. \,\, 
\end{align}
where $\digamma(t)=\frac{1}{2} \left(  P_k(t) - P_k^{ave} \right)^2 
+ q_k[t-1] \left(  P_k(t) - P_k^{ave} \right)$ and $\mathrm{MSE}_k(P_k(t))=
\left( | \mathbf{b}^H(t) \hat{\mathbf{h}}_k(t)|\sqrt{P_k(t)} - 1 \right)^2 +P_k(t) \sigma_h^2 \left\| \mathbf{b}(t) \right\|^2$. The projected gradient descent (PGD) method is used to derive the numerical solution for $(\mathcal{P}1.3)$. Specifically, the power coefficient $P_k(t)$ is iteratively updated as $P^{q+1}_k(t)=P^{q}_k(t)-\delta\nabla_{P^{q}_k(t)}$, where $\nabla_{P^{q}_k(t)}$ represents the gradient of the objective function $\digamma(t)+V\mathrm{MSE}_k(P_k(t)) $ with respective to $P_k(t)$ on the $q$-th iteration, and can be written as 
$\nabla_{P^{q}_k(t)}=\big(P_k^q(t)-P_k^{\mathrm{ave}}\big)
  + q_k[t-1]
  + V\!\left(| \mathbf{b}^H(t) \hat{\mathbf{h}}_k(t)|^2-\frac{| \mathbf{b}^H(t) \hat{\mathbf{h}}_k(t)|}{\sqrt{P_k^q(t)}}+\sigma_h^2 \left\| \mathbf{b}(t) \right\|^2\right)$ and $\delta$ is the step size for the the gradient descent method  \cite{Lya,Lya1}.
After each gradient descent, the power coefficient needs to be projected into the feasible region to guarantee that the maximum transmission power limit is met by
\begin{equation} \label{queue11}
    P_k^*(t) =
\begin{cases}
P_k^{\max}, & \text{if } P^*_k(t) > P_k^{\max},\\
P_k^*(t), & \text{if } P^*_k(t) \leq P_k^{\max}.
\end{cases}
\end{equation}

\begin{algorithm}[t]
\centering
\begin{minipage}{0.45\textwidth}  
\caption{Causal Imperfect CSI aided Optimization for Transmit Coefficients Design in OTA-FL}
\label{Alg2}

\textbf{Input:} Channel information for the current round $\hat{\boldsymbol{h}}_k(t)$; stopping condition $\epsilon$, maximum iteration time $Q$\;

\For{$k \in \{1,2, ..., K\}$}{
    \Repeat{$\{\digamma(t)+V\overline{\mathrm{MSE}}_k(P_k(t))\}^{q+1} - \{\digamma(t)+V\overline{\mathrm{MSE}}_k(P_k(t))\}^{q} \leq \epsilon$ or $q=Q$}{
        (i): Construct the objective function via \eqref{obj1.3};\\
        (ii): Calculate the gradient of the objective function $\Delta L_k(t)+V\overline{\mathrm{MSE}}_k(P_k(t))$ with respect to $P_k(t)$, and update the transmit coefficient $P_k(t)$ based on $P_k^{q+1}(t) = P_k^q(t) - \delta \Delta_{P_k^q(t)}$;\\
        (iii): Update $P_k^{q+1}(t)$ according to \eqref{queue11};\\
        (iv): $q = q + 1$;
    }
}

\textbf{Output:} $P_k^*(t)$ ;
\end{minipage}
\end{algorithm}

The detailed procedure for solving problem~\((\mathcal{P}1.3)\) is summarized in Algorithm~2. From Algorithm~2, it can be observed that the causal CSI, i.e., only the present round's CSI is needed for computing the transmit coefficients. Up to this stage, the joint optimization of the transmit coefficients and the receive combining vector has been completed to minimize the long-term MSE in OTA-FL systems with imperfect CSI. In particular, two novel schemes are developed for the transmit coefficient design under non-causal and causal CSI, respectively. In the following section, we will provide numerical simulations to demonstrate the effectiveness of the proposed methods.

\begin{algorithm}[t]
\centering
\begin{minipage}{0.45\textwidth}  
\caption{Overall Solution for Problem $(\mathcal{P}0)$}
\label{Alg3}

\textbf{Input}: Channel information; stopping condition $\epsilon_0$, maximum iteration time $Q_0$\;

\textbf{Initialization:} The transmit power at each local device, receive combining vector, and iteration index $q = 0$\;

    \Repeat{$\dfrac{\overline{\mathrm{MSE}}_{q-1} - \overline{\mathrm{MSE}}_q}{\overline{\mathrm{MSE}}_q} \leq \epsilon_0$ or $q = Q_0$}
    {
    (i): $ q = q + 1$;\\
        (ii): Given the receive combining vector $\{\boldsymbol{b}(t)\}_{q-1}$, based on the input channel information, if it is non-causal information $\{\hat{\boldsymbol{h}}_k(t)\}_{t=0}^{T-1}$, optimize $\{P_k(t)\}_q$ according to Algorithm 1; if it is causal information $\hat{\boldsymbol{h}}_k(t)$, optimize  $\{P_k(t)\}_q$ according to Algorithm 2;\\
        (iii):  Given $\{P_k(t)\}_{q}$, update $\{\boldsymbol{b}(t)\}_q$ via \eqref{b(t)};\\
    }

\textbf{Output:}$\{P_k^*(t)\}$ and $\{\boldsymbol{b}^*(t)\}$.
\end{minipage}
\end{algorithm}

\subsection{Complexity Analysis}
Algorithm~3 summarizes the overall solution to $(\mathcal{P}0)$, where the transmit coefficients are updated by Algorithm~1 under non-causal CSI and by Algorithm~2 under causal CSI. 
For Algorithm~1, the condition check in \eqref{power_condi} and the closed-form update in \eqref{power1} have complexity $\mathcal{O}(KT)$. When the bisection-based update in \eqref{power2} is required, the complexity becomes $\mathcal{O}(KT\log(1/\varepsilon))$, with $\varepsilon$ denoting the bisection accuracy. The receive combining update in Theorem~3 is dominated by constructing $K$ outer products of $M$-dimensional vectors and inverting an $M\times M$ matrix, yielding $\mathcal{O}(T(KM^2+M^3))$ over $T$ rounds. For Algorithm~2, each round requires $\mathcal{O}(KM)$ for effective-channel computation and $\mathcal{O}(KQ)$ for $Q$ PGD iterations, resulting in a total complexity of $\mathcal{O}(TK(M+Q))$.

\section{Performance Evaluation}
This section provides numerical experiments to assess the efficacy and capabilities of our developed approaches. The corresponding simulation setup is described as follows. The wireless links between local devices and the PS are assumed to follow IID Rayleigh fading across aggregation rounds~\cite{channel_model}. Unless otherwise specified, the number of local devices is set to \(K = 20\). For each device, the transmission signal-to-noise ratio (SNR) is expressed as 
\(\text{SNR}_k = P_k^{\text{ave}} / \sigma_0^2\), which is set to a random value between 10~dB to 15~dB~\cite{Yinan} to simulate the heterogeneous power budgets. The maximum transmission power is then configured as \(P_k^{\max} = 2P_k^{\text{ave}}\) for each user.
For the causal imperfect CSI-based transmit coefficient design (Algorithm~2), we set \(V=10\) as an MSE-oriented operating point. This choice places greater emphasis on aggregation MSE minimization, while the virtual queue mechanism maintains the long-term average
transmit power constraint~\cite{Lya,Lyaa}. For the initial value of the virtual queue for local device $k$, we assume $c_k$ is randomly chosen from the interval $[0,\,0.5]$ to represent heterogeneous initial queue states across the devices.
The FL-related simulation parameters are summarized as follows.

\begin{figure*}[t!]
    \centering
	\begin{minipage}[b]{0.43\linewidth}
		\centering		\includegraphics[width=\linewidth,trim=10 8 10 8,clip]{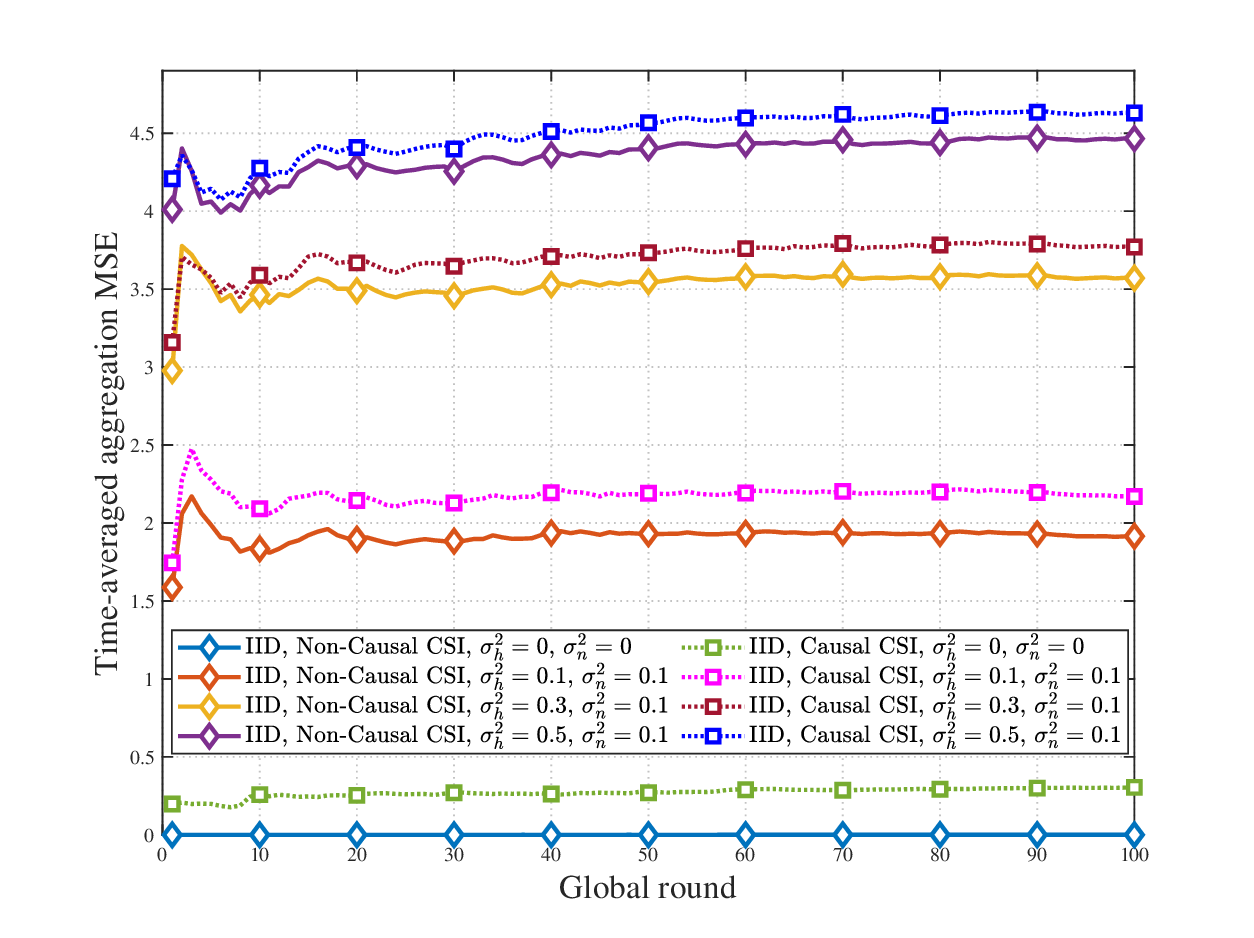}
		\footnotesize{{(a) Time-averaged MSE on Fashion-MNIST using IID data.}}
		\label{IID_MSE}
	\end{minipage}
	\begin{minipage}[b]{0.43\linewidth}
		\centering
		\includegraphics[width=\linewidth,trim=10 8 10 8,clip]{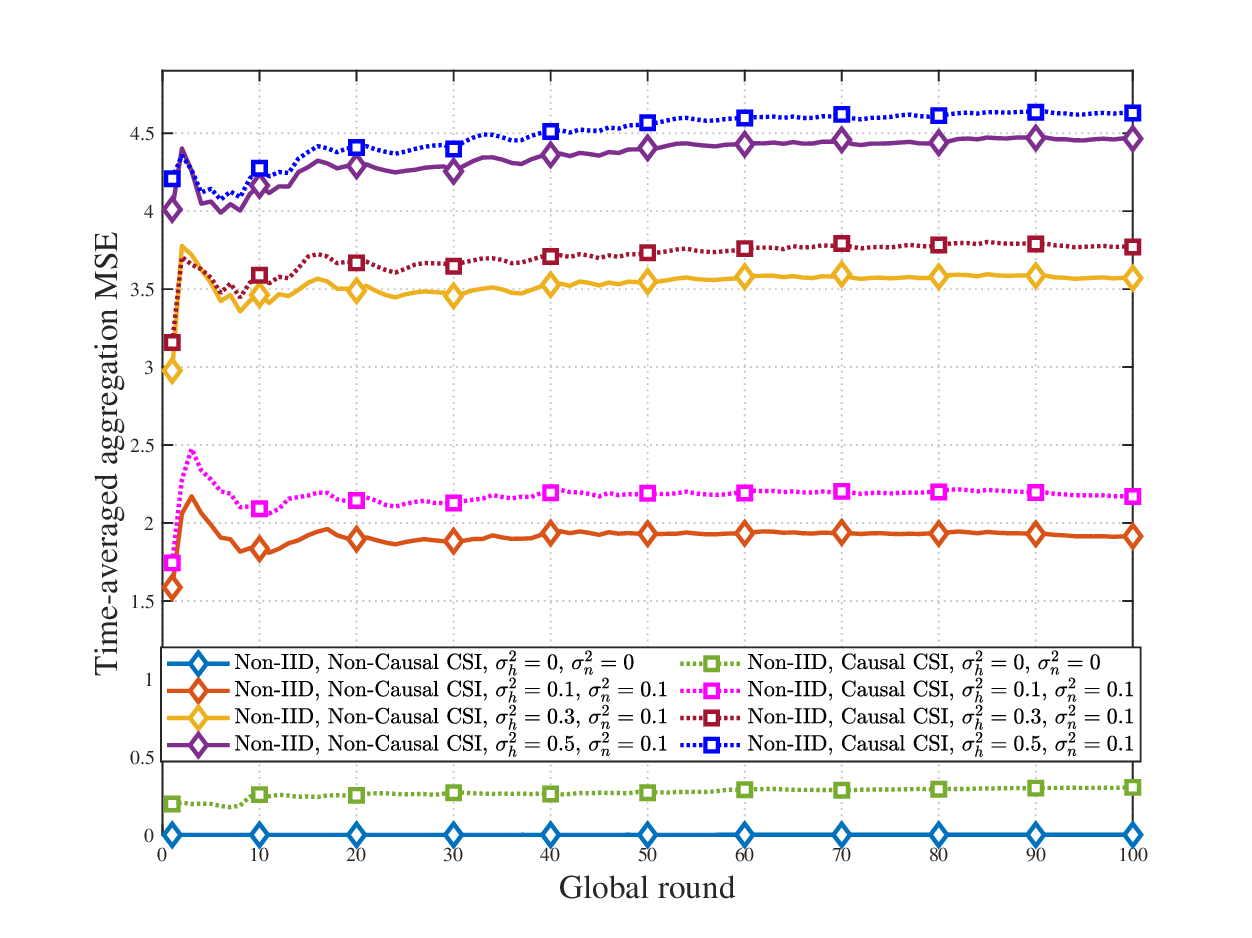}
		\footnotesize{{(b) Time-averaged MSE on Fashion-MNIST using non-IID data.}}
		\label{NonIID_MSE}
	\end{minipage}
    
	\centering
	\begin{minipage}[b]{0.43\linewidth}
		\centering		\includegraphics[width=\linewidth,trim=10 8 10 8,clip]{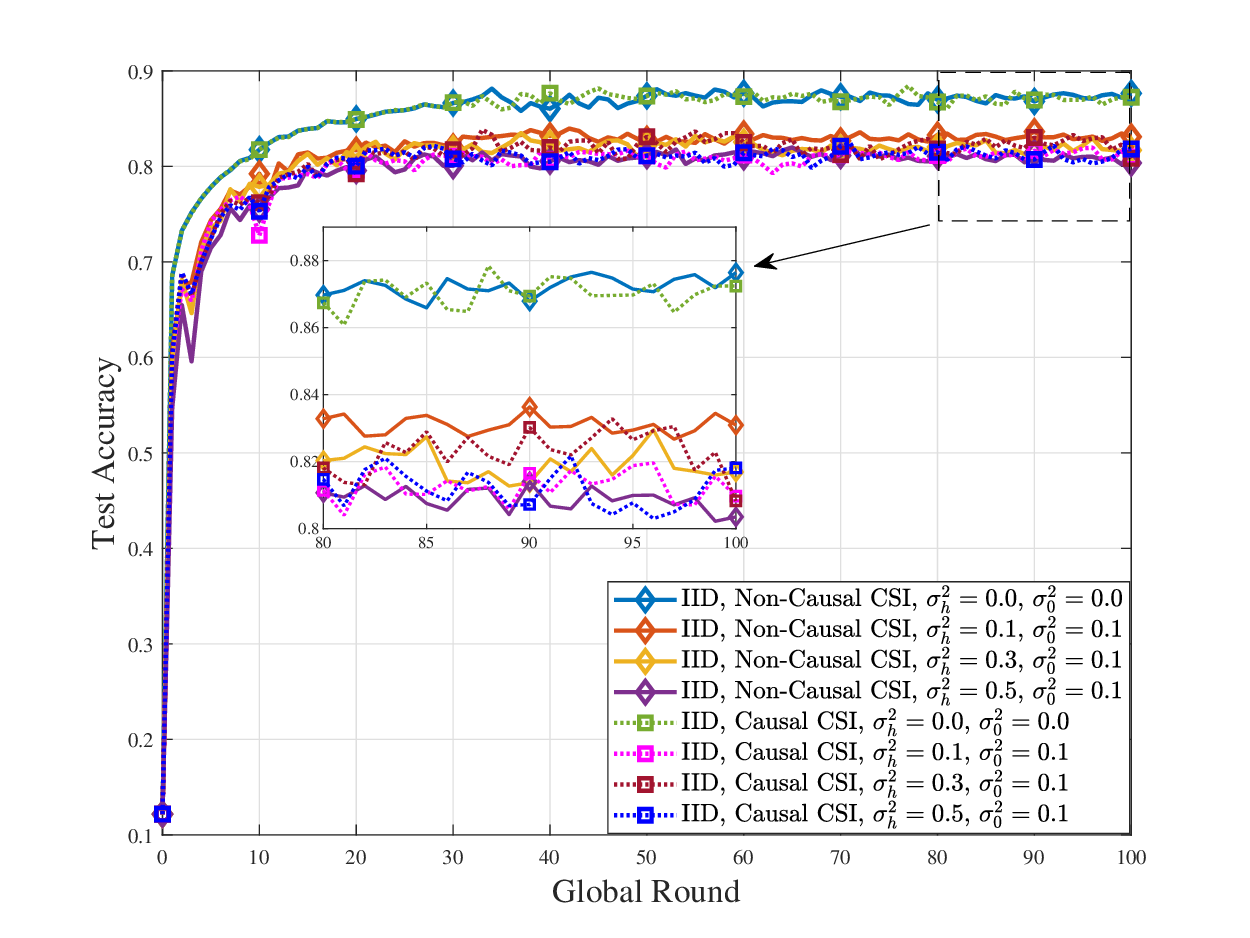}
		\footnotesize{{(c) Test accuracy on Fashion-MNIST using IID training data.}}
		\label{IID_MNIST}
	\end{minipage}
	\begin{minipage}[b]{0.43\linewidth}
		\centering
		\includegraphics[width=\linewidth,trim=10 8 10 8,clip]{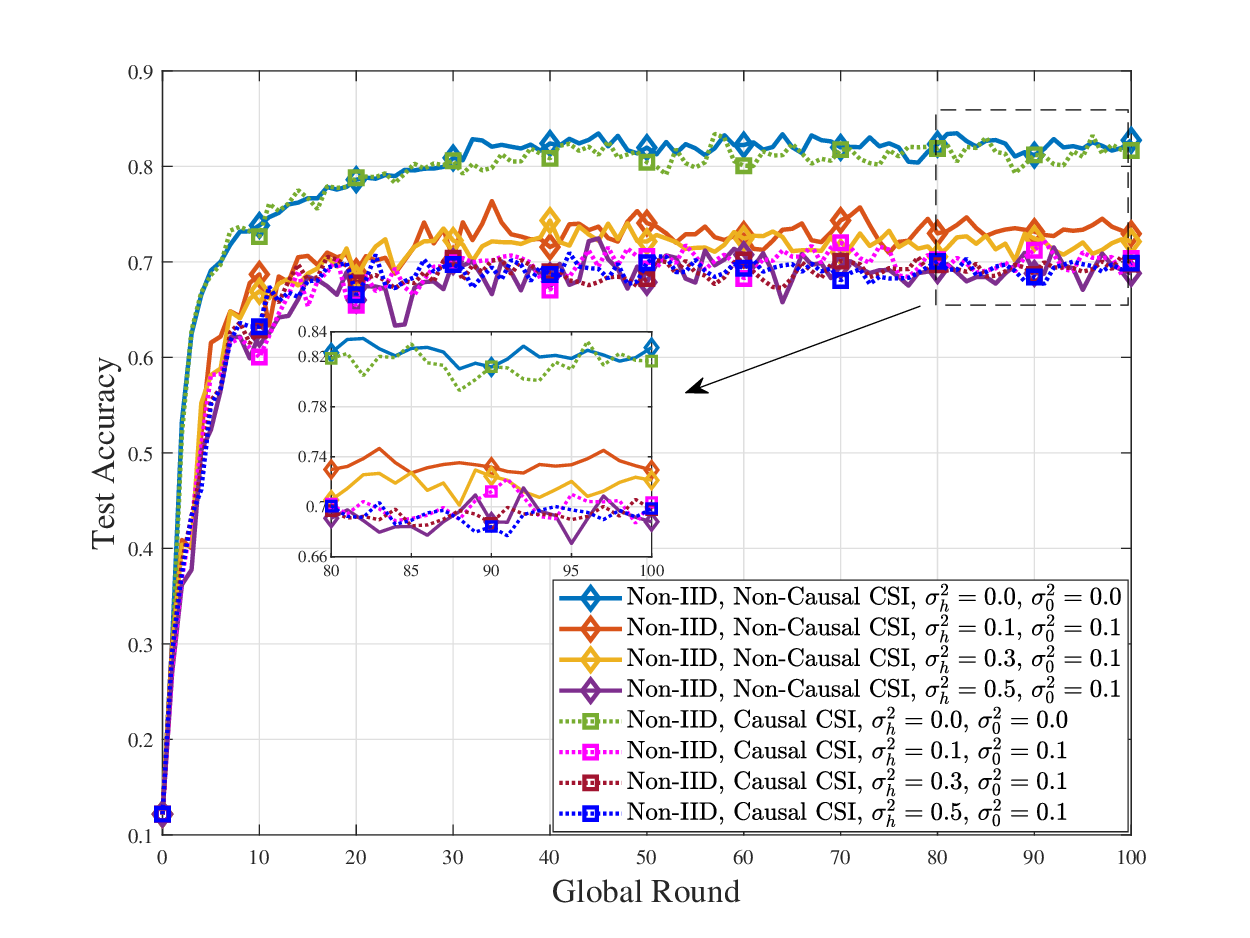}
		\footnotesize{{(d) Test accuracy on Fashion-MNIST using non-IID training data.}}
		\label{NonIID_MNIST}
	\end{minipage}
    \captionsetup{justification=justified, width=0.98\textwidth}  
 \caption{{Time-averaged MSE and test accuracy of the proposed non-causal CSI and causal CSI aided optimization methods on Fashion-MNIST dataset using IID training data and non-IID training data, respectively. }}
 \label{MNIST}
 \vspace{-1.2em}
\end{figure*}

\begin{figure*}[t!]
	\centering
	\begin{minipage}[b]{0.43\linewidth}
		\centering
		\includegraphics[width=\linewidth,trim=10 8 10 8,clip]{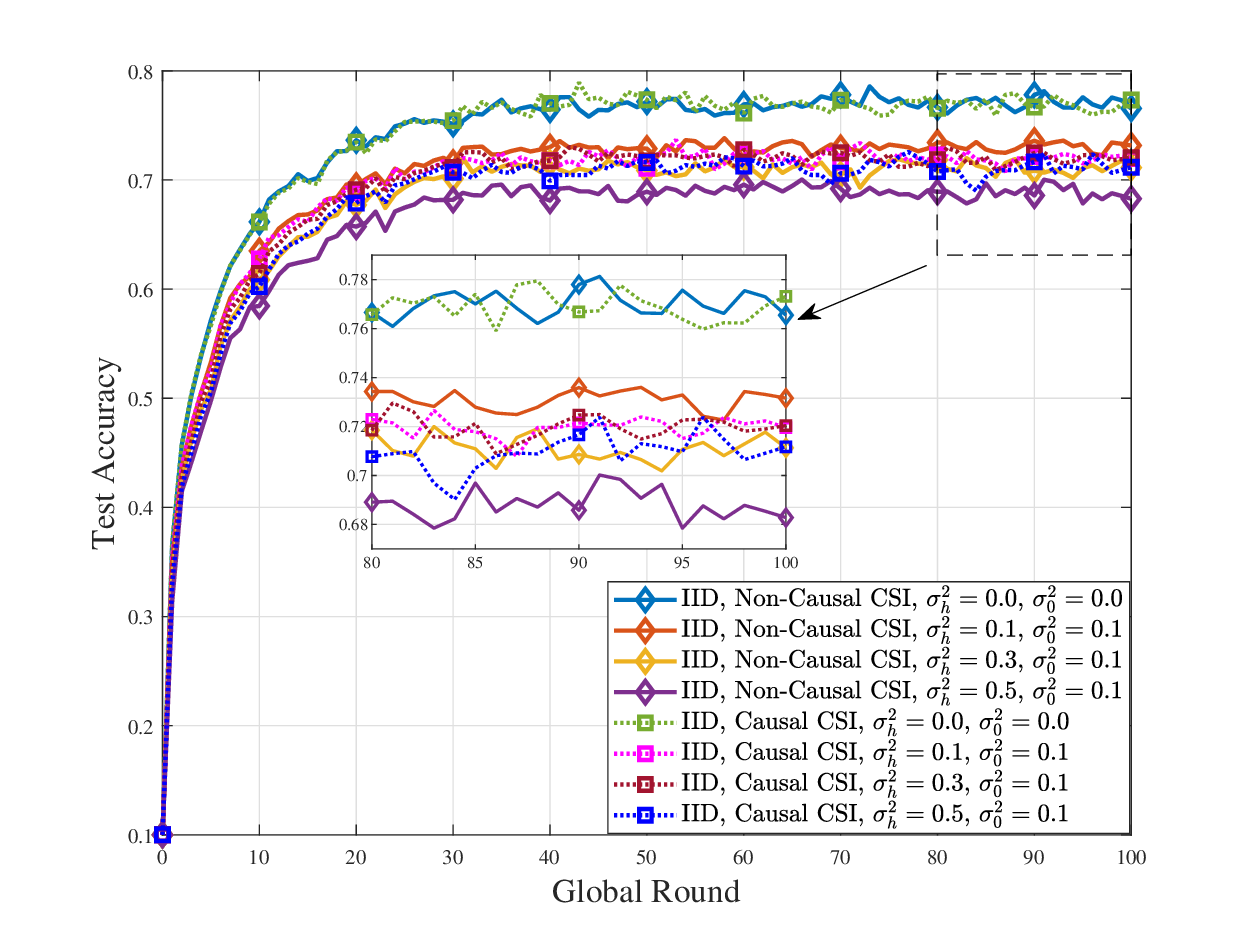}
		\footnotesize{{(a) Test accuracy on CIFAR-10 using IID training data.}}
		\label{IID_CIFAR}
	\end{minipage}
	\begin{minipage}[b]{0.43\linewidth}
		\centering
		\includegraphics[width=\linewidth,trim=10 8 10 8,clip]{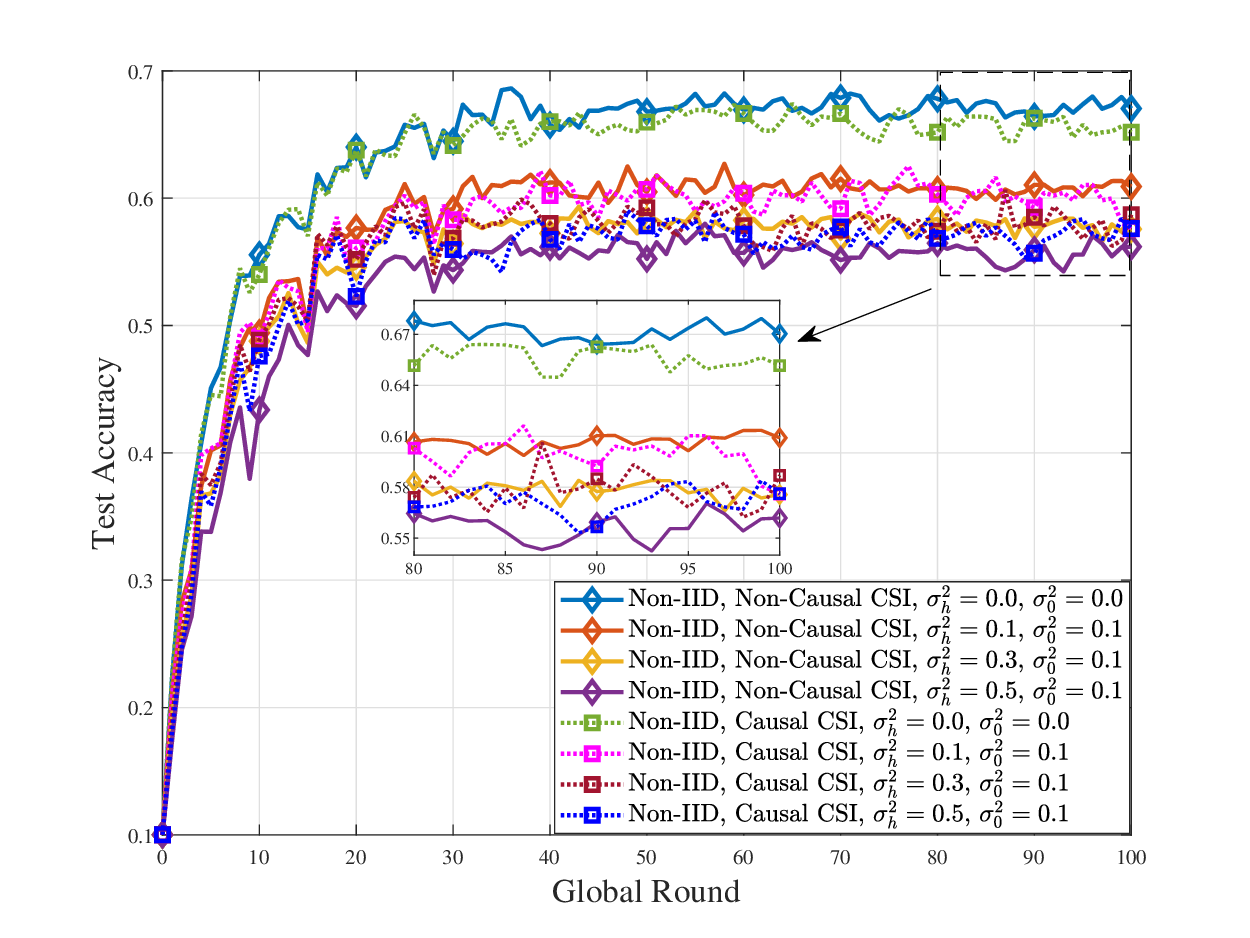}
		\footnotesize{{(b) Test accuracy on CIFAR-10 using non-IID training data.}}
		\label{NonIID_CIFAR}
	\end{minipage}
    \captionsetup{justification=justified, width=0.98\textwidth}  
 \caption{{Test accuracy of the proposed non-causal CSI and causal CSI aided optimization methods on CIFAR-10 dataset.}}
 \label{CIFAR}
 \vspace{-1.2em}
\end{figure*}

\begin{itemize}
 \item \textit{Datasets}: We use Fashion-MNIST, CIFAR-10, and CIFAR-100 as standard OTA-FL benchmarks~\cite{normalize0,Assumption,dataset0}.
Both IID and non-IID settings are evaluated. In the IID case, we set $\sigma_d=0$ in \eqref{OBJ0}, randomly shuffle the training data, and assign 5 out of 200 evenly divided shards to each device. In the non-IID case, the data are first sorted by class labels and then divided into 200 shards, with each device randomly assigned 5 shards~\cite{Yinan}.

\item \textit{FL Neural Network:} For the Fashion-MNIST dataset, we employ a CNN architecture consisting of two convolutional layers followed by two fully connected layers.
For CIFAR-10, we adopt a deeper CNN model with three convolutional layers and two fully connected layers to accommodate the increased data complexity. For CIFAR-100, we utilize ResNet-18, a widely adopted deep convolutional neural network composed of 18 layers, including convolutional, batch normalization, and fully connected layers, which is well-suited for large-scale image classification tasks.

\end{itemize}

\subsection{Performance Evaluation of the Proposed Algorithm}

\begin{figure*}[t!]
	\centering
	\begin{minipage}[b]{0.43\linewidth}
		\centering
		\includegraphics[width=\linewidth,trim=10 8 10 8,clip]{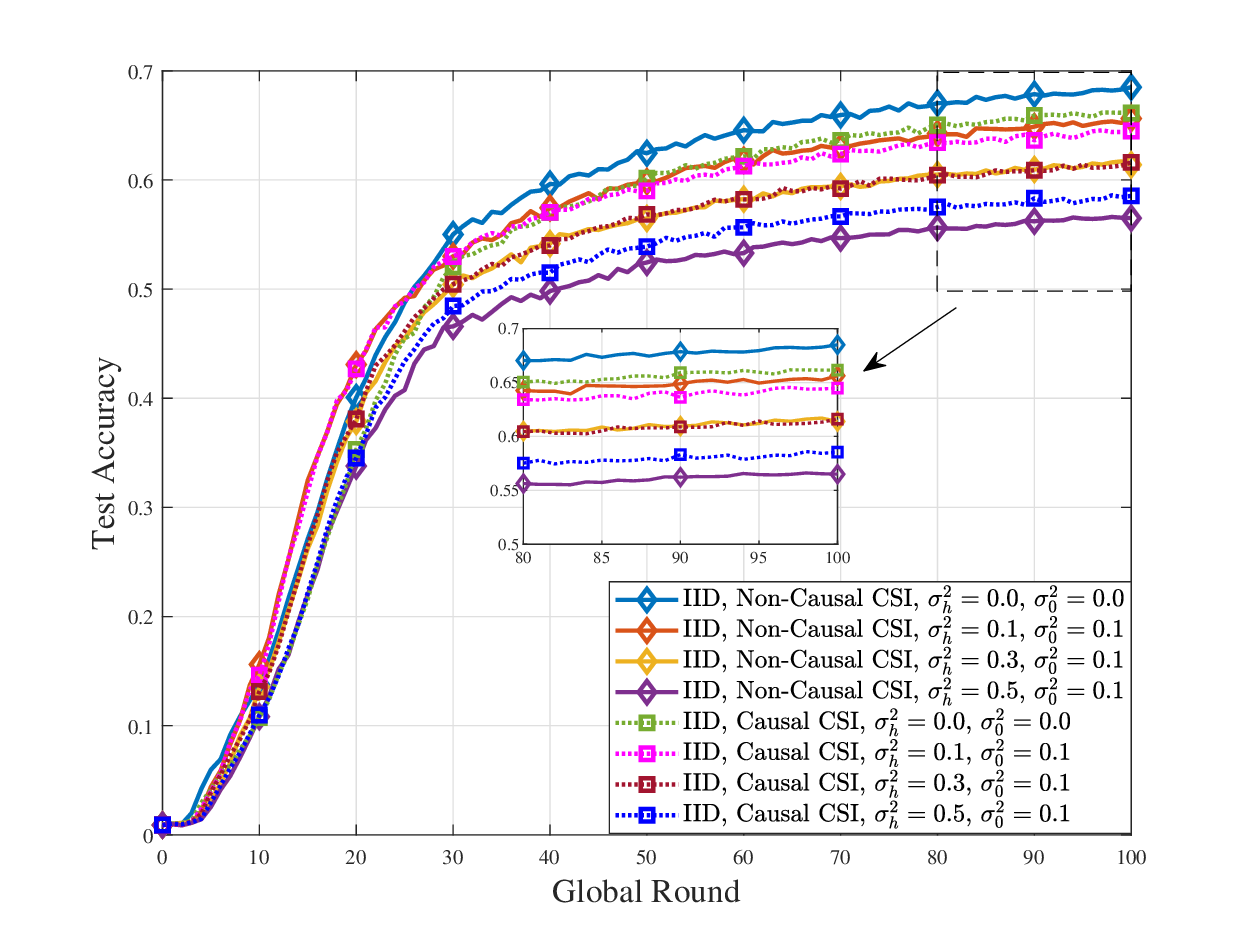}
		\footnotesize{(a) Test accuracy on CIFAR-100 using IID training data.}
		\label{IID_CIFAR100}
	\end{minipage}
	\begin{minipage}[b]{0.43\linewidth}
		\centering
		\includegraphics[width=\linewidth,trim=10 8 10 8,clip]{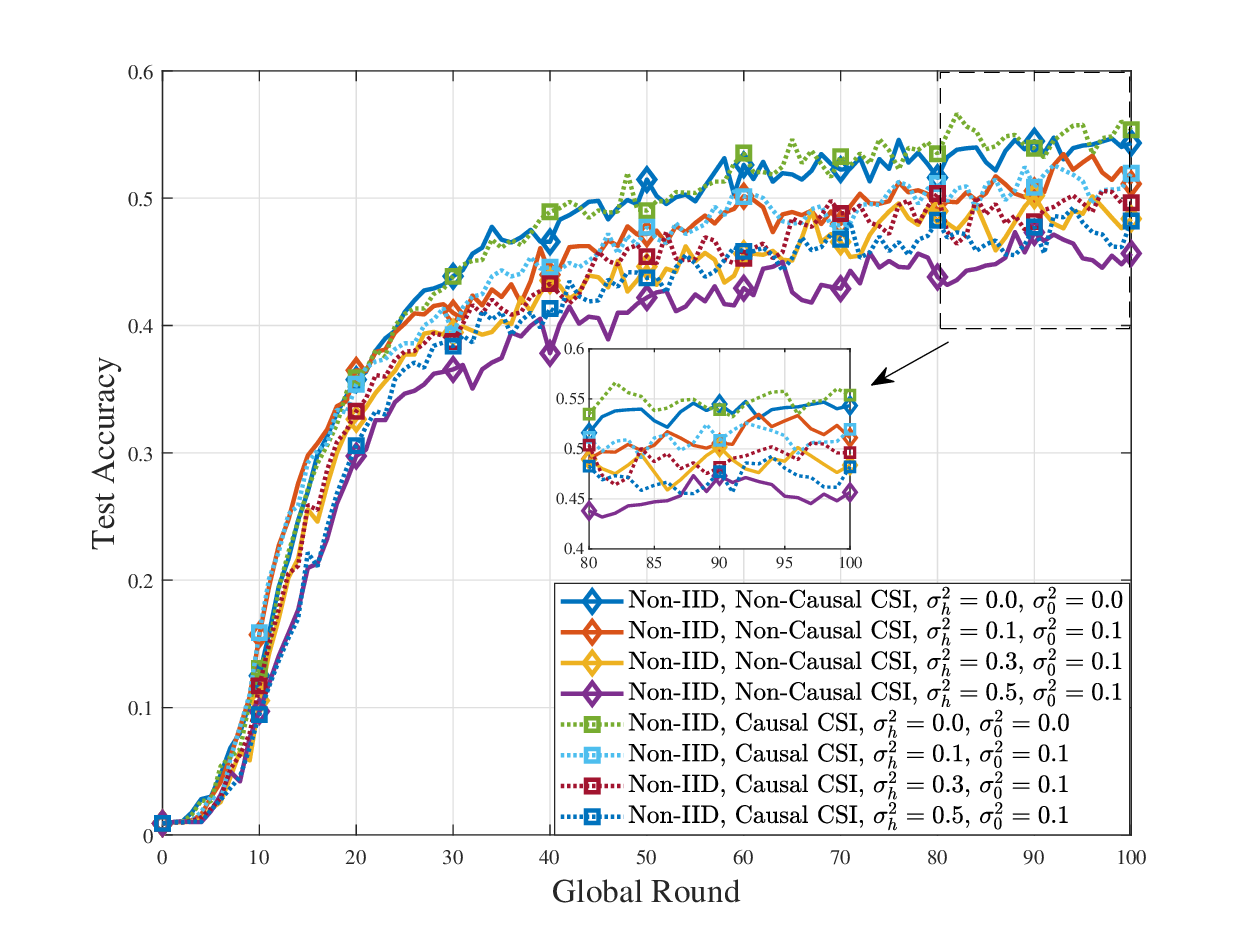}
		\footnotesize{(b) Test accuracy on CIFAR-100 using non-IID training data.}
		\label{NonIID_CIFAR100}
	\end{minipage}
    \captionsetup{justification=justified, width=0.98\textwidth}  
 \caption{Test accuracy of the proposed non-causal CSI and causal CSI aided optimization methods on CIFAR-100 dataset.}
 \label{CIFAR100}
\end{figure*}

\begin{figure*}[t!]
\vspace{-1.2em}
	\centering
	\begin{minipage}[b]{0.43\linewidth}
		\centering
		\includegraphics[width=\linewidth]{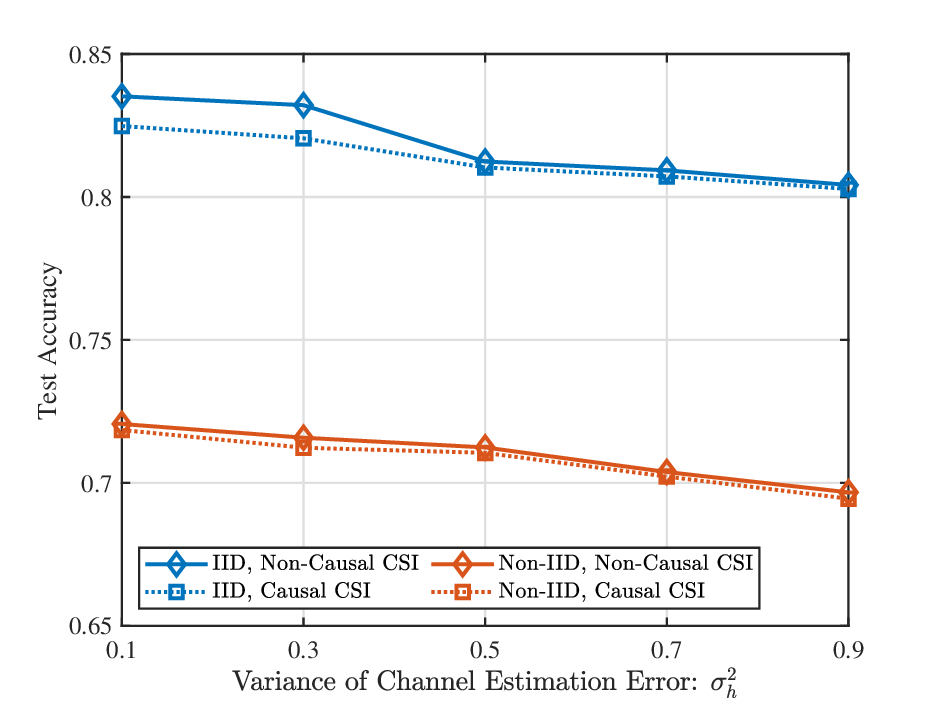}
		\footnotesize{{(a) Test accuracy after 100 rounds of global model aggregation on Fashion-MNIST dataset with different channel estimation errors.}}
		\label{IID_CIFAR}
	\end{minipage}
	\begin{minipage}[b]{0.43\linewidth}
		\centering
		\includegraphics[width=\linewidth]{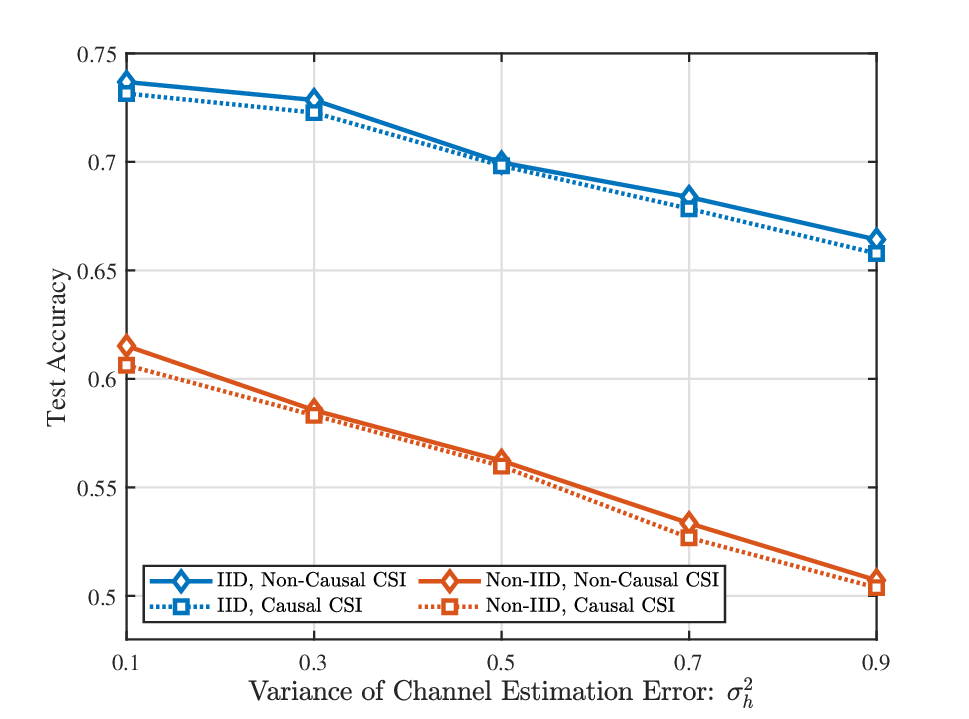}
		\footnotesize{{(b) Test accuracy after 100 rounds of global model aggregation on CIFAR-10 dataset with different channel estimation errors.}}
		\label{NonIID_CIFAR}
	\end{minipage}
    \captionsetup{justification=justified, width=0.98\textwidth}  
 \caption{{Test accuracy after 100 global aggregation rounds for the proposed non-causal CSI and causal CSI based optimization methods on Fashion-MNIST and CIFAR-10 datasets.}}
 \label{Final_Acc}
\end{figure*}

To verify the performance of the proposed non-causal CSI and causal-CSI aided optimization methods, we first test them on the Fashion-MNIST, CIFAR-10 and CIFAR-100 datasets using both IID and non-IID training data.  In both Fig. \ref{MNIST}, Fig. \ref{CIFAR} and Fig. \ref{CIFAR100} , the $\sigma_h^2=0$, $\sigma_0^2=0$ case serves as the upper bound in OTA-FL, where local parameters go through perfect uploading channels with no channel estimation error and background noise. To show the relationship of MSE minimization and test accuracy optimization, in Fig. \ref{MNIST}, we plot the time-averaged MSE and corresponding test accuracy under both IID training data and non-IID training data for Fashion-MNIST. Specifically, subfigure~(a) and subfigure~(b) report the time-averaged aggregation MSE, which is the direct optimization goal of the objective function $(\mathcal{P}1)$, whereas subfigure~(c) and subfigure~(d) report the corresponding test accuracy. From Fig.~\ref{MNIST} (a) and Fig.~\ref{MNIST} (b), we can see that the time-averaged aggregation MSE depends on the wireless channels, channel-estimation errors, transmit coefficients,
and receive combining vector, the IID and non-IID cases have the same MSE curves since we use the the same wireless channel realizations. Nevertheless, their test accuracies differ substantially because of the different training data distributions and model training dynamics. Moreover, as the channel estimation error $\sigma_h^2$ increases, the aggregation MSE increases systematically, whereas the corresponding accuracy degradation depends strongly on the data distribution and training dynamics. This observation demonstrates that although a larger aggregation MSE generally tends to degrade learning performance, aggregation MSE and test accuracy do not exhibit a
one-to-one or strictly monotonic correspondence. These results confirm that the aggregation MSE directly quantifies
the communication distortion introduced during OTA aggregation, whereas test accuracy evaluates the resulting end-to-end learning performance after such distortion propagates through successive global model updates. In particular, test accuracy is jointly influenced by the aggregation error, data heterogeneity, local model optimization,
and stochastic training dynamics. Consequently, the same aggregation MSE can result in different test accuracies under IID and non-IID data, while different MSE values do not necessarily lead to proportional changes in test accuracy. Therefore, aggregation MSE measures communication fidelity, whereas test accuracy reflects the ultimate task-level utility of the learned model. Based on this observation, we will focus on test accuracy in the following simulations. 
From Fig. \ref{MNIST} - Fig. \ref{CIFAR100}, we can also observe that the performance gaps between the upper bound and the proposed designs with imperfect channels are quite small in both IID and non-IID scenarios, which verifies that our proposed design can effectively manage the channel estimation error and noise for OTA-FL global model aggregation process. Besides, comparing the IID training dataset (Fig. \ref{MNIST} (a), Fig. \ref{CIFAR} (a) and Fig. \ref{CIFAR100} (a)) with the non-IID training dataset (Fig. \ref{MNIST} (b), Fig. \ref{CIFAR} (b) and Fig. \ref{CIFAR100} (b)), we can observe that the test accuracy with the non-IID dataset suffers with more severe fluctuation since the local gradients point to different directions with the non-IID training dataset. Besides, for both IID and non-IID results trained on Fashion-MNIST, CIFAR-10 and CIFAR-100 datasets, we can observe that the proposed non-causal CSI and causal-CSI aided schemes have almost the similar performance, which verifies the effectiveness of the proposed Lyapunov optimization based system design.

To further evaluate the effectiveness of the proposed non-causal CSI and causal CSI aided imperfect channel management schemes for OTA-FL, we present the final test accuracies after 100 rounds of global model aggregation under both IID and non-IID training datasets for Fashion-MNIST and CIFAR-10 datasets. Note that we focus on these two datasets as they represent learning tasks with different levels of complexity, i.e., grayscale and simple level versus colored and moderate-scale image classification tasks, which are sufficient to demonstrate the effectiveness and robustness of the proposed schemes. Similar performance trends were also observed on CIFAR-100 as shown before, thus are omitted for brevity. The results in Fig.~\ref{Final_Acc} show that the test accuracy generally decreases as the channel estimation error variance increases, particularly under non-IID training, confirming the adverse effect of CSI uncertainty on OTA-FL. Nevertheless, the causal scheme remains close to the non-causal benchmark across all considered channel estimation error levels. These results indicate that the proposed causal design can maintain competitive learning performance using only current-round CSI under imperfect-CSI conditions.
\begin{table}[t]
\centering
\caption{Running time of the proposed schemes on different computing platforms.}
\label{tab:runtime_platforms}
\begin{tabular}{llccc}
\toprule
Platform & Scheme
& Mean per-round time
& End-to-end time \\
\midrule
Tesla V100S
& Non-causal
& $20.432 $
& $2050.5 $ \\

Tesla V100S
& Causal
& $20.371 $
& $2041.1$ \\
\midrule
NVIDIA H100
& Non-causal
& $11.765 $
& $1181.9 $ \\

NVIDIA H100
& Causal
& $11.816 $
& $1184.1$ \\
\bottomrule
\end{tabular}

\vspace{1mm}

\parbox{\linewidth}{%
\scriptsize
\justifying
\setlength{\parindent}{0pt}
\textit{Note:} All runtime is reported in seconds. The mean per-round time is calculated over rounds 2--100, with round 1 excluded from the average to mitigate one-time startup effects. The end-to-end time denotes the total time from model and data-handler initialization to the completion of round 100.
}
\end{table}

\begin{table}[t!]
\centering
\caption{{Transmit power, virtual queue, and aggregation MSE for non-causal and causal schemes.}}
\label{tab:tight_power_results}

\scriptsize
\setlength{\tabcolsep}{2pt}
\renewcommand{\arraystretch}{1.12}

\begin{adjustbox}{max width=\columnwidth}
\begin{tabular}{@{}lccccc@{}}
\toprule
Scheme
& \shortstack{Avg. transmit\\power}
& \shortstack{Mean power\\utilization}
& \shortstack{Active-device\\fraction (\%)}
& \shortstack{Mean queue\\backlog}
& \shortstack{Time-averaged\\MSE} \\
\midrule

Non-causal
& $0.798$
& $0.868$
& $53.50$
& N/A
& $1.968$ \\

Causal
& $0.813$
& $0.875$
& $51.00$
& $0.344$
& $1.991$ \\

\bottomrule
\end{tabular}
\end{adjustbox}
\begin{minipage}{0.95\linewidth}
\scriptsize
\textit{Note:} Mean power utilization
is defined as $K^{-1}\sum_{k=1}^{K}
\bar{P}_k/P_k^{\mathrm{ave}}$, where
$\bar{P}_k=T^{-1}\sum_{t=1}^{T}P_k^*(t)$.
A device is active if
$\bar{P}_{k}/P_{k}^{\mathrm{ave}}\geq 0.9$.
The mean queue backlog is averaged over devices and communication rounds.  All reported statistics are averaged over 20 independent channel realizations. Queue-related entry is N/A for non-causal scheme since it does not have a Lyapunov virtual queue.
\end{minipage}
\end{table}
To show the actual computational cost of the proposed methods, we run experiments on two heterogeneous computing platforms: (i) Tesla V100S, and (ii) NVIDIA H100. For fair comparisons, both platforms used one GPU, 14 CPU threads, the same CIFAR-10 non-IID data partition. We separately measured the per-round runtime, and end-to-end runtime. The first global round was treated as a warm-up and excluded from the calculation
of the per-round mean, while it remained included in the end-to-end runtime. As summarized in Table~\ref{tab:runtime_platforms}, over 100 global rounds, the two schemes exhibit comparable mean per-round runtime on each platform, requiring approximately 20.4 seconds on the Tesla V100S and 11.8 seconds on the NVIDIA H100. The NVIDIA H100 platform achieves an approximately \(1.73\times\) speedup over the Tesla V100S platform, reducing the end-to-end runtime by approximately \(42\%\) for both schemes.

To explicitly examine whether the long-term average power constraints are active, we report the average transmit power, virtual queue, and aggregation MSE statistics in Table~\ref{tab:tight_power_results}. The average transmit powers (averaged over devices and communication rounds) of the non-causal and causal schemes are $0.798$ and $0.813$, respectively. The corresponding mean power utilizations are
$0.868$ and $0.875$, respectively, where power
utilization is defined as the ratio between the actual average
transmit power of each device and its prescribed long-term
average transmit power limit. Thus, on average, the non-causal and causal
schemes consume $86.8\%$ and $87.5\%$ of their respective
device-specific power budgets. Moreover, $53.50\%$ and $51.00\%$ of the devices under the non-causal and causal schemes, respectively, utilize at
least $90\%$ of their individual power budgets. These results show
that the similarity between the two schemes is not caused by inactive
power constraints, since a substantial fraction of devices operate
close to their power budget boundaries. For the causal scheme, the
mean virtual-queue backlog is $0.344$. This nonzero backlog
indicates that the average power constraint actively influences the
online transmission decisions. Finally, the time-averaged MSEs
of the non-causal and causal schemes are $1.968$ and
$1.991$, respectively, which shows that the causal scheme increases the time-averaged MSE by only approximately $1.17\%$ compared to the non-causal benchmark. Therefore, in the considered scenarios,
the per-round CSI together with the virtual queue state provides sufficient information for effective per-round power control, while
non-causal knowledge of future CSI offers only a marginal MSE improvement.

\begin{figure}[t]
\vspace{-1em}
    \centering
    \includegraphics[width=\linewidth]{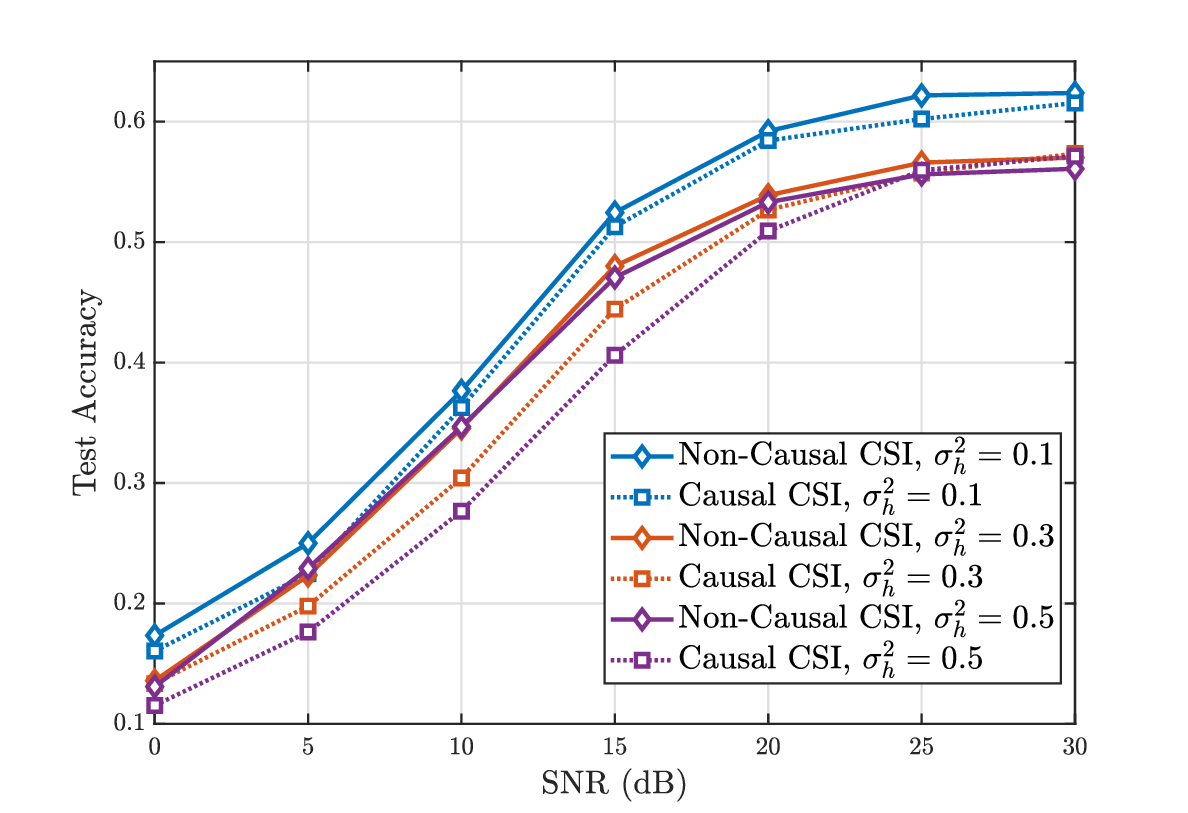}
    \caption{Test accuracy versus different SNR values with CIFAR-10 non-IID training data.}
    \label{fig:SNR}
\end{figure}

\begin{figure} [t]
\vspace{-0.8em}
    \centering
    \includegraphics[width=\linewidth]{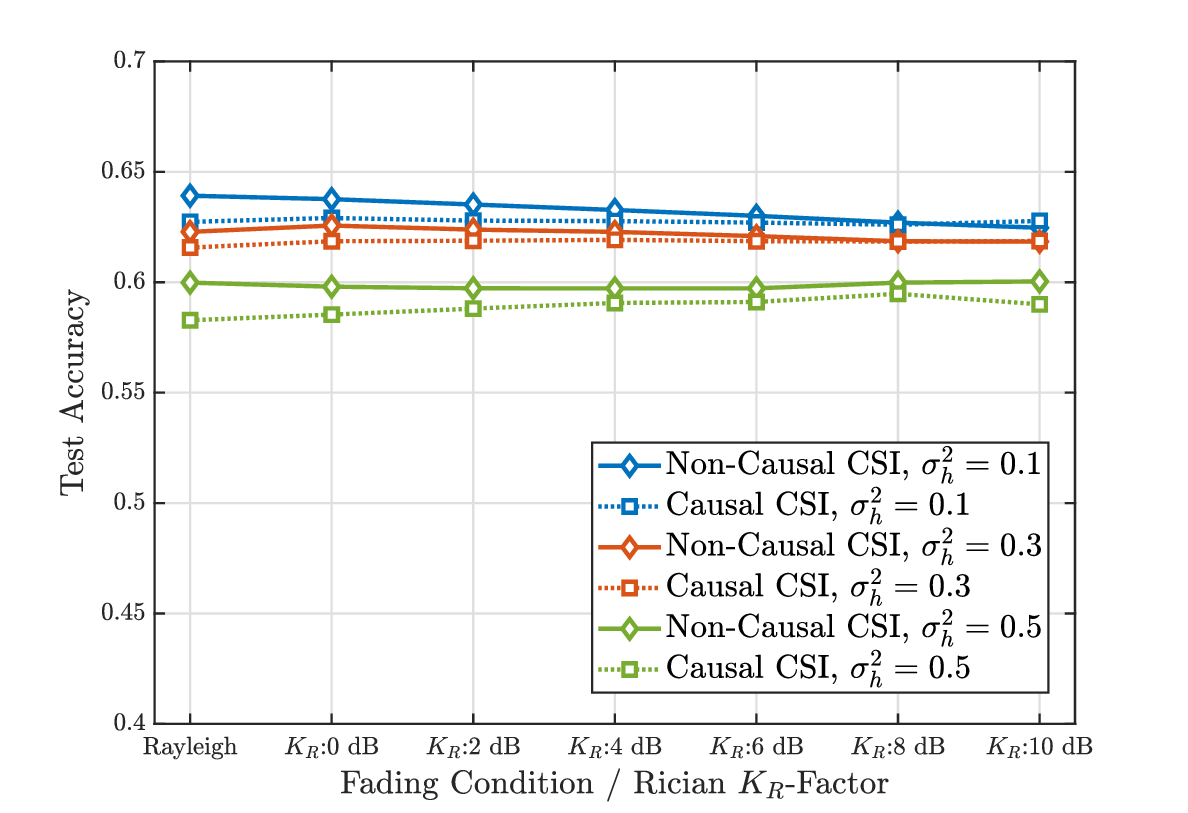}
    \caption{Test accuracy under different channel realizations with CIFAR-10 non-IID training data.}
    \label{fig:channel}
\end{figure}

To further show the robustness of the proposed schemes, we further evaluate them over a wider SNR range from 0 dB to 30 dB as shown in Fig. \ref{fig:SNR} and under different channel realizations as shown in Fig. \ref{fig:channel}. Specifically, Fig.~\ref{fig:SNR} evaluates the test accuracy of the proposed non-causal and
causal CSI-aided schemes over a wide SNR range under different channel estimation errors. In this simulation, the noise power is set as $\sigma_0^2=0.1$. As the SNR increases, the test accuracy of all schemes improves significantly, since the reduced receiver noise leads to more accurate OTA aggregation. Meanwhile, for a given SNR, a smaller channel estimation error variance consistently yields better learning performance, demonstrating the adverse impact of CSI uncertainty on OTA-FL. It can also be observed that the performance gain from increasing the SNR gradually diminishes in the high-SNR regime. This is because, when the receiver noise becomes sufficiently small,
the aggregation distortion caused by imperfect CSI becomes the dominant performance-limiting factor, which restricts the further increase of test accuracy. Fig.~\ref{fig:channel} further evaluates the proposed schemes under different channel realizations, including Rayleigh fading and Rician fading with
$K_R$-factors ranging from $0$ to $10$ dB. To ensure a fair comparison, the Rician channels are normalized as
$\mathbf{h}
=
\sqrt{{K_R}/({K_R+1})}\mathbf{h}_{\mathrm{LoS}}
+
\sqrt{{1}/({K_R+1})}\mathbf{h}_{\mathrm{NLoS}}$,
such that the average channel power remains unchanged for different $K_R$-factors. Therefore, varying $K_R$ mainly changes the relative contributions of the line-of-sight (LoS) and non-LoS (NLoS) components, rather than the overall channel gain. As shown in Fig.~7, the test accuracies of both the non-causal and causal CSI-aided schemes remain nearly unchanged when the channel varies from Rayleigh fading to Rician fading with different $K_R$ factors. These results indicate that the proposed schemes do not rely on a specific fading realization, and can maintain stable learning performance under diverse wireless propagation conditions.
\begin{figure*}[t]
	\centering
	\begin{minipage}[b]{0.43\linewidth}
		\centering		\includegraphics[width=\linewidth,trim=10 8 10 8,clip]{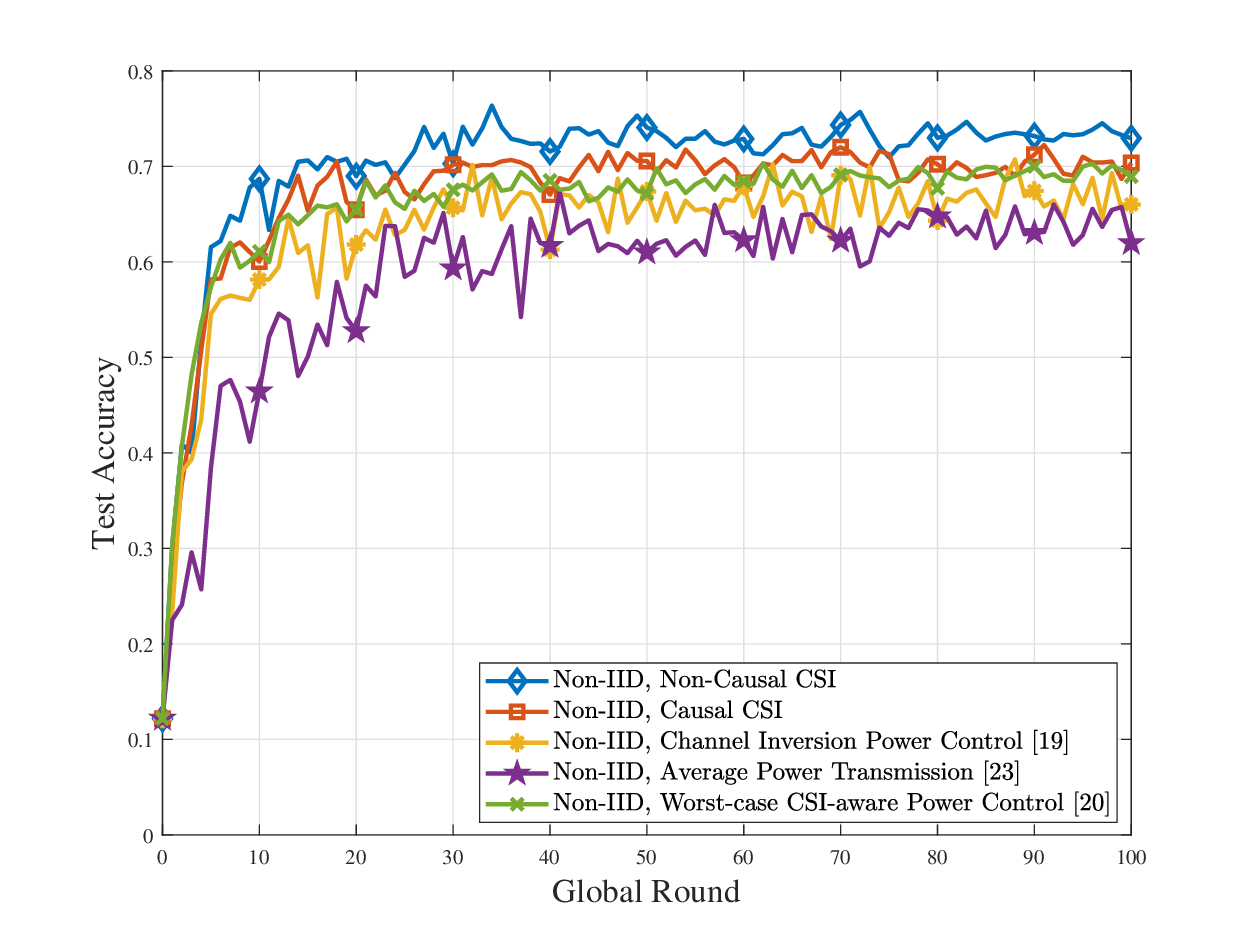}
		\footnotesize{(a) Performance comparison of different transmit coefficient designs with non-IID training data on the Fashion-MNIST dataset.}
		\label{ComparisonIIDMNIST}
	\end{minipage}
	\begin{minipage}[b]{0.43\linewidth}
		\centering
		\includegraphics[width=\linewidth,trim=10 8 10 8,clip]{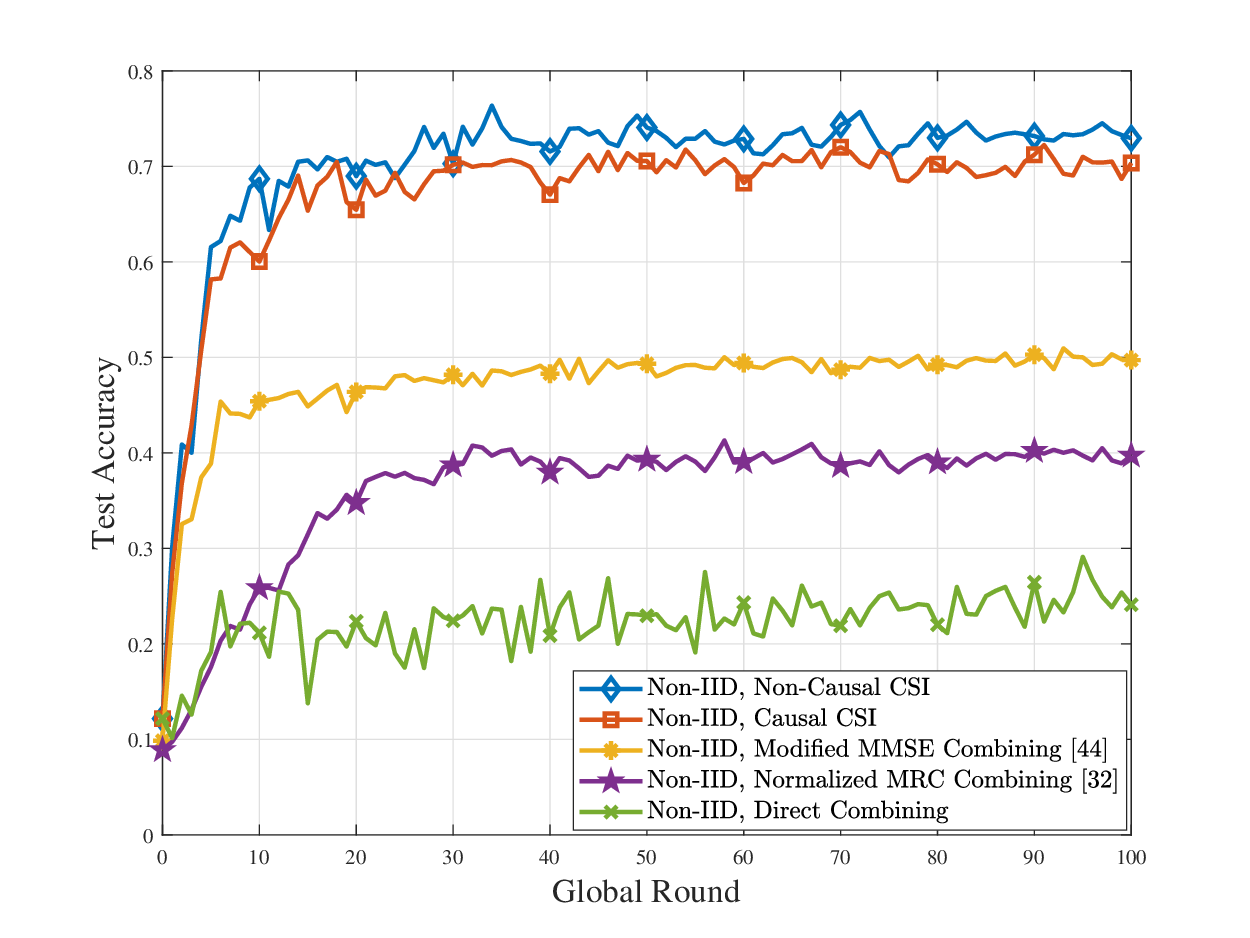}
		\footnotesize{(b) Performance comparison of different receive combining vector designs with non-IID training data on the Fashion-MNIST dataset.}
		\label{ComparisonNonIIDMNIST}
	\end{minipage}
    \captionsetup{justification=justified, width=0.98\textwidth}  
 \caption{Test accuracy of the proposed schemes and the comparison schemes on Fashion-MNIST dataset when $\sigma_h^2=0.1$ and $\sigma_0^2=0.1$ with non-IID training data.}
 \label{ComparisonMNIST}
  \vspace{-1.2em}
\end{figure*}
\begin{figure*}[h!]
	\centering
	\begin{minipage}[b]{0.43\linewidth}
		\centering
		\includegraphics[width=\linewidth,trim=10 8 10 8,clip]{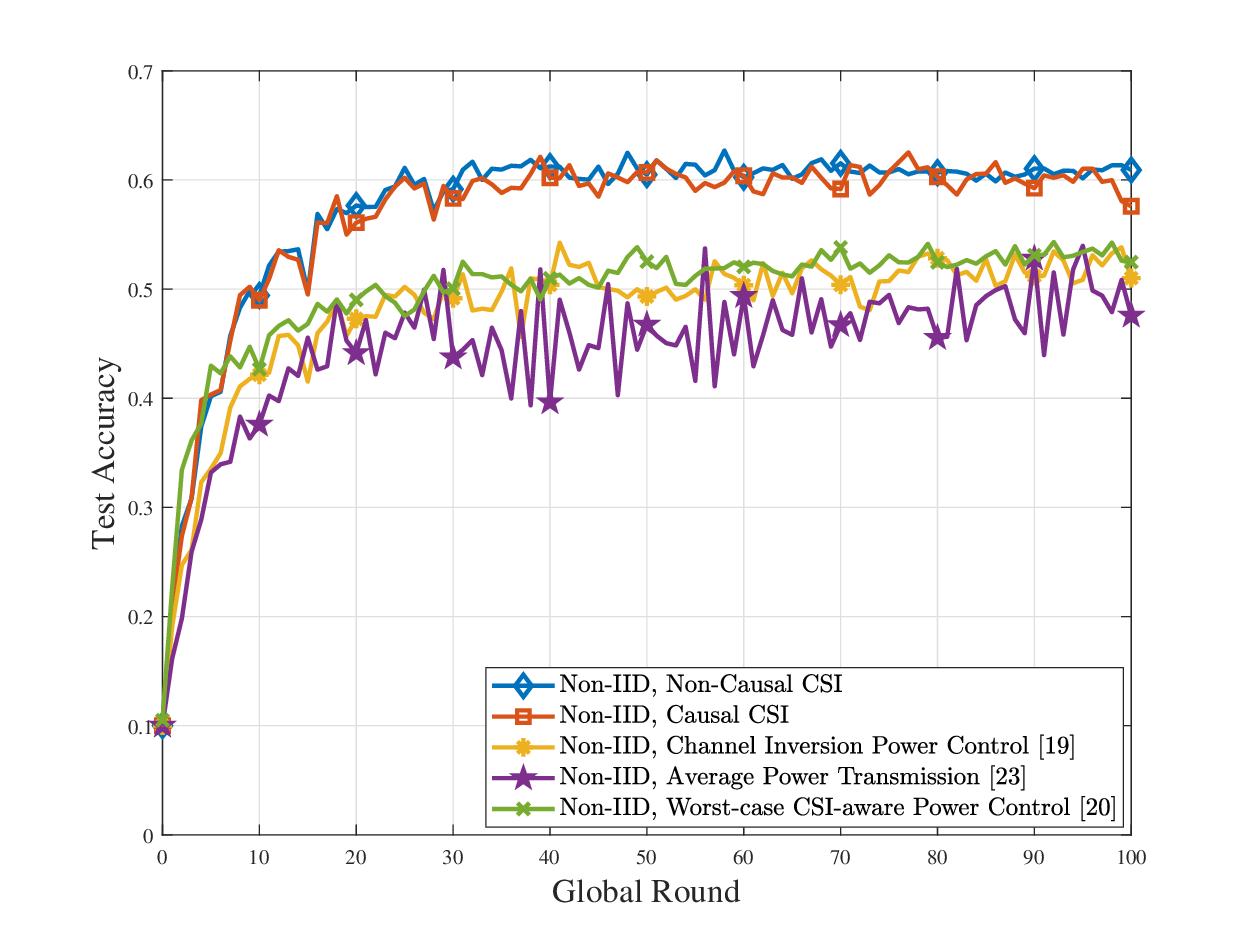}
		\footnotesize{(a) Performance comparison of different transmit coefficient designs with non-IID training data on the CIFAR-10 dataset .}
		\label{ComparisonIIDCIFAR}
	\end{minipage}
	\begin{minipage}[b]{0.43\linewidth}
		\centering
		\includegraphics[width=\linewidth,trim=10 8 10 8,clip]{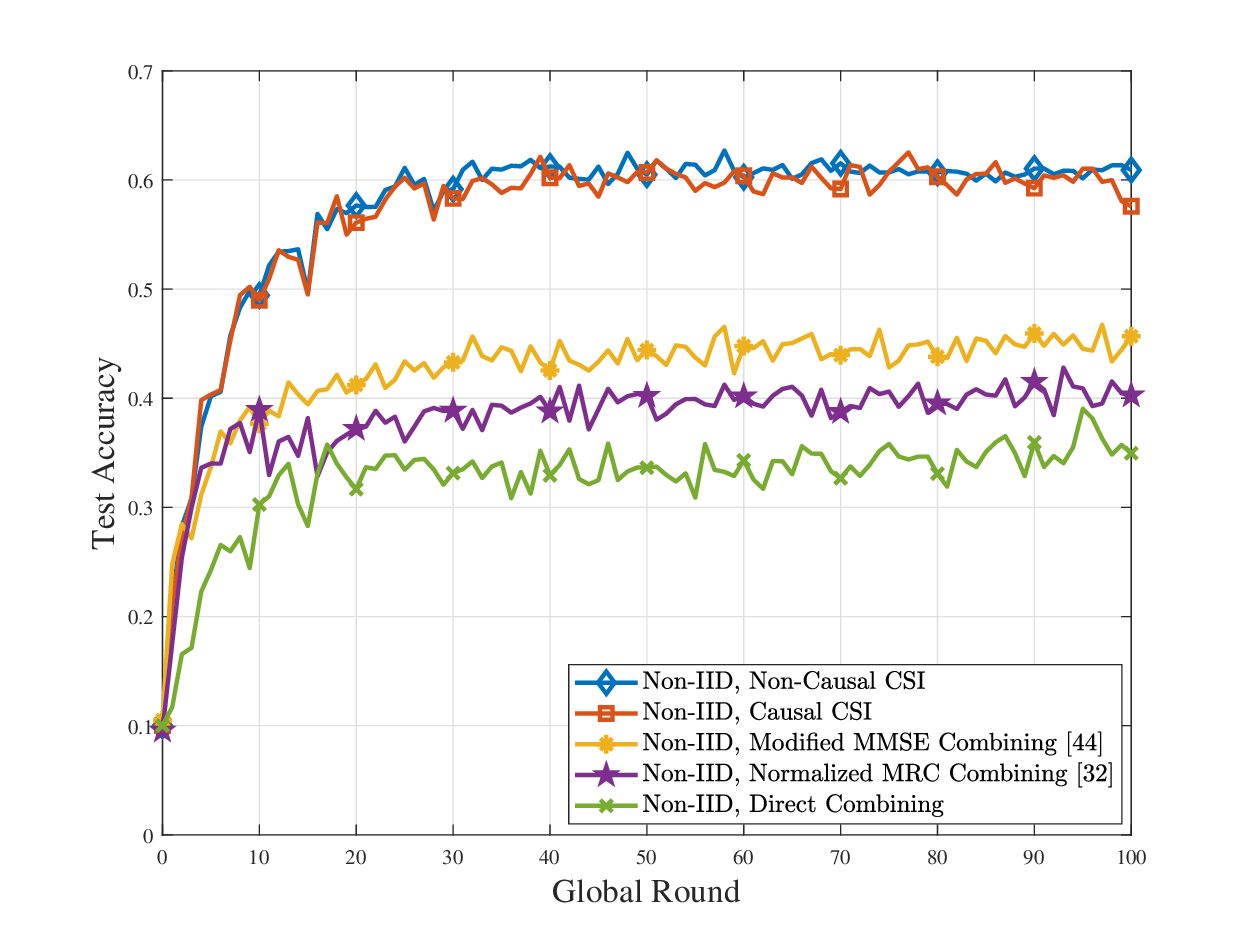}
		\footnotesize{(b) Performance comparison of different receive combining vector designs with non-IID training data on the CIFAR-10 dataset.}
		\label{ComparisonNonIIDCIFAR}
	\end{minipage}
    \captionsetup{justification=justified, width=0.98\textwidth}  
 \caption{Test accuracy of the proposed schemes and the comparison schemes on the CIFAR-10 dataset when $\sigma_h^2=0.1$ and $\sigma_0^2=0.1$ with non-IID training data.}
 \label{ComparisonCIFAR}
\end{figure*}

\begin{figure*}[t]
 \vspace{-0.3cm}
	\centering
	\begin{minipage}[b]{0.43\linewidth}
		\centering		\includegraphics[width=\linewidth,trim=10 8 10 8,clip]{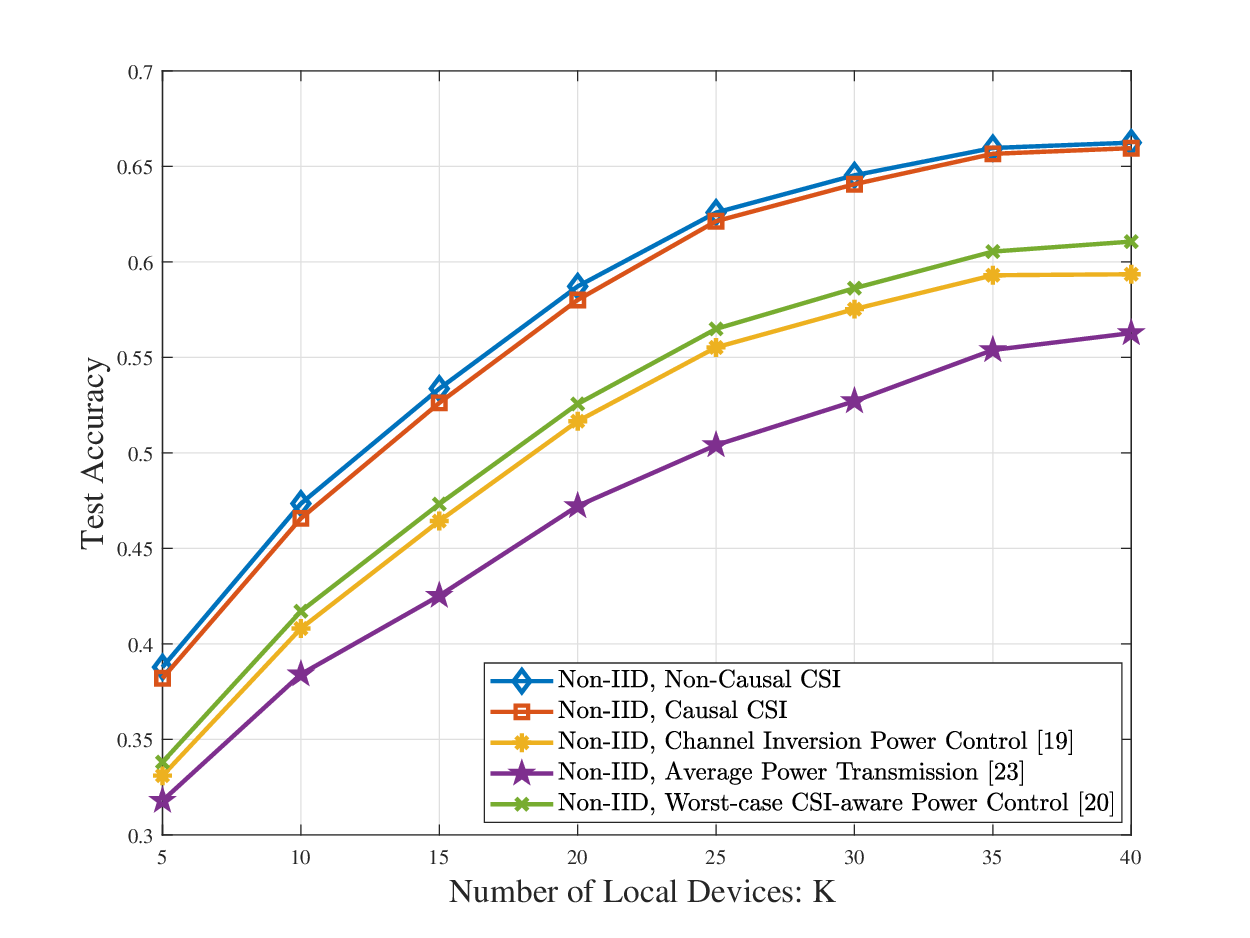}
		\footnotesize{(a) Performance comparison of different transmit coefficient designs when the number of local devices increases.}
		\label{TransNumInc}
	\end{minipage}
	\begin{minipage}[b]{0.43\linewidth}
		\centering
		\includegraphics[width=\linewidth,trim=10 8 10 8,clip]{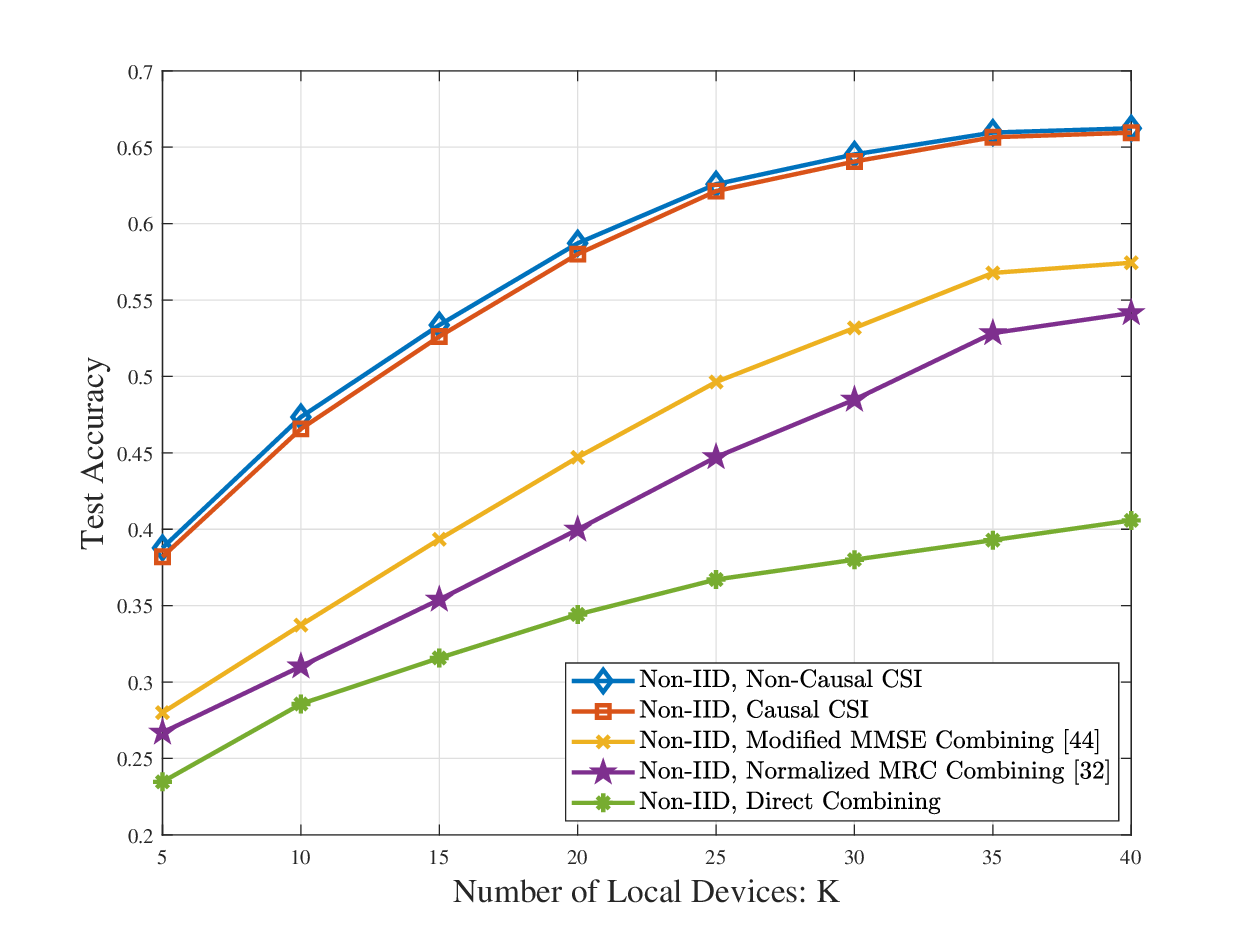}
		\footnotesize{(b)
        Performance comparison of different receive combining vector designs when the number of local devices increases.}
		\label{RecNumInc}
	\end{minipage}
    \captionsetup{justification=justified, width=0.98\textwidth}  
 \caption{Test accuracy when the number of local devices increases for the CIFAR-10 dataset with non-IID training data.}
 \label{numberinc}
 \vspace{-1.4em}
\end{figure*}

\subsection{Performance Comparison with Benchmark Schemes}
To further demonstrate the effectiveness of the proposed methods, we focus on the more challenging scenario with non-IID training dataset, since our previous results have verified that the proposed schemes perform comparably well on both IID and non-IID training data. In this section, we specifically compare the proposed approaches against the following baseline methods:
\begin{itemize}
    \item \textbf{Average power transmission:} Each edge device transmits with the average transmit power $P^{ave}_k$ and the aligned phase \cite{Yinan}, i.e., $\mu_k(t)=\sqrt{P^{ave}_k}\frac{ \hat{\boldsymbol{h}}^H_k(t)\boldsymbol{b}(t)}{|\boldsymbol{b}^H(t)\hat{\boldsymbol{h}}_k(t)|}$.
    \item \textbf{Channel inversion power control:} Each local device sets its transmit coefficient based on channel inversion, i.e., $\mu_k(t) = \sqrt{P^{ave}_k} \frac{\min_{i \in \mathcal{K}} \|\hat{\boldsymbol{h}}_i(t)\|}{\|\hat{\boldsymbol{h}}_k(t)\|} \frac{\hat{\boldsymbol{h}}_k^H(t) \boldsymbol{b}(t)}{|\boldsymbol{b}^H(t)\hat{\boldsymbol{h}}_k(t)|}$, where the local device with the poorest channel uses up its transmit power \cite{Related1}.

    \item \textbf{Worst-case CSI-aware power control:} Each local device is classified into either a full-power group or a channel-inversion group. Specifically, the devices with the $i^*$ weakest effective channels $|\boldsymbol{b}^H(t)\hat{\boldsymbol{h}}_k(t)|$ transmit at full power (i.e., $\sqrt{P^{ave}_k}$), while the remaining devices apply channel inversion under a worst-case phase difference approximation, i.e., $\mu_k(t) = \frac{|\mathbf{b}^H(t) \hat{\mathbf{h}}_k(t)|}{a^* (|\mathbf{b}^H(t) \hat{\mathbf{h}}_k(t)|^2 + \sigma_h^2)}  \frac{\hat{\mathbf{h}}_k^H(t) \mathbf{b}(t)}{|\mathbf{b}^H(t) \hat{\mathbf{h}}_k(t)|}$, where $a^*$ is the optimal receiver scaling factor and $\sigma_h^2$ is the channel estimation error variance. The  optimal receiver scaling factor $a^*$ and the critical number $i^*$ are selected to minimize the MSE under the worst-case phase misalignment assumption while guarantee the long-term average power constraint considered in this paper \cite{Related2}.

     \item \textbf{Normalized MRC combining:} The receive combining vector at the PS is the normalized variant of maximum ratio combining (MRC), where each channel vector is scaled by its squared norm to balance the contribution of different local devices \cite{MRC_Com}, i.e., $\boldsymbol{b}(t)=\sum_{k=1}^K \frac{\hat{\boldsymbol{h}}_k^H(t)}{||\hat{\boldsymbol{h}}_k(t)||^2}$.

     \item \textbf{Modified MMSE combining:} The receive combining vector at the PS is designed based on the minimum mean-square error (MMSE) criterion in \cite{Com3}, which minimizes the upper bound of the one-round aggregation MSE under the transmit power constraint. This design focuses solely on minimizing noise amplification, without accounting for channel estimation errors in the combining vector design.

    \item \textbf{Direct combining:} We set each element in the receive combining vector to $1$, which means we do not apply any optimization method to the received signals.
\end{itemize}
During the comparison, only the corresponding transmit coefficients or receive combining design is replaced, while the other one still follows the proposed design. For example, when employing normalized MRC combining or direct combining at the PS side, the device side still adopts the proposed non-causal CSI aided transmit coefficients design scheme for fair comparison. We first present the test accuracies of our proposed schemes and the comparison schemes in Fig. \ref{ComparisonMNIST} and Fig. \ref{ComparisonCIFAR} for Fashion-MNIST and CIFAR-10 datasets, respectively. From the results, we can see that our proposed non-causal CSI and causal CSI schemes can achieve the best performances in both IID and non-IID cases on both Fashion-MNIST and CIFAR-10 datasets. For the transmit coefficients design in Fig. \ref{ComparisonMNIST} (a) and Fig. \ref{ComparisonCIFAR} (a), we can see that the channel inversion power control, average power transmission and worst-case CSI-aware power control can achieve comparable performance compared with our proposed schemes. But for the combining vector disigns shown in Fig. \ref{ComparisonMNIST} (b) and Fig. \ref{ComparisonCIFAR} (b), we can observe that direct combining and normalized MRC combining perform poorly since they do not have the ability to mitigate the global model aggregation distortion caused by imperfect CSI. Modified MMSE combining also achieves inferior performance compared with the proposed schemes, as it only minimizes the one-round aggregation MSE and does not account for channel estimation errors.
Besides, compared to direct combining design, the MRC strategy achieves better performance as it leverages the conjugate of the channel to enhance signal alignment. In contrast, direct combining treats all signals from all local devices equally without considering for any channel variations, resulting in the worst performance.

Fig.~\ref{numberinc} presents the training performance of our developed methods and baseline approaches as a function of the number of local devices using the CIFAR-10 dataset.
Results from Fig.~\ref{numberinc}  confirm that the learning performance scales with network size due to enhanced data diversity.
In particular, the test accuracy grows rapidly when \(K \leq 20\), whereas the improvement slows down for \(K \geq 25\). This trend arises because excessive participation of local devices introduces data redundancy during the OTA-FL training process.
Also, we can see that our proposed non-causal CSI and causal CSI aided schemes can always achieve better test accuracy than other benchmark schemes.

\section{Conclusion} \label{sec:conclusion}
In this paper, we studied the long-term MSE minimization problem for OTA-FL systems under imperfect CSI conditions. Based on the convergence analysis, we first derived the accumulated MSE over multiple global aggregation rounds. We then proposed a joint optimization strategy for transmit-receive parameters to minimize the long-term MSE bound. Since the proposed algorithm requires non-causal CSI, we then introduced the concept of virtual queues to represent the cumulative energy consumption, which successfully decouples the time-based coupling, so that causal CSI can be used in each round for transmit coefficients design. Finally, numerical evaluations and comparisons verified that our proposed scheme can effectively manage the imperfect OTA aggregation channels.

\vspace{-0.5 em}
\appendices
\begingroup
\setlength{\abovedisplayskip}{3pt plus 1pt minus 1pt}
\setlength{\belowdisplayskip}{3pt plus 1pt minus 1pt}
\setlength{\abovedisplayshortskip}{2pt plus 1pt minus 1pt}
\setlength{\belowdisplayshortskip}{2pt plus 1pt minus 1pt}
\setlength{\jot}{1pt}
\section{Proof of Theorem 1} \label{Theorem1}
To assist in the derivation and verification of Theorem~1, we first present Lemmas~1 – 3, whose detailed proofs can be referred to~\cite{Yinan}.

\textit{Lemma 1}: Under Assumptions~2 and~3, the following inequality holds:
\setlength{\jot}{1pt}
\begin{align}
&- \lambda \mathbb{E} \!\left[ \langle \nabla F(\boldsymbol{w}(t)), \boldsymbol{\theta}(t) \rangle \right]\nonumber \le 
  \! \!- \frac{\lambda I}{2} \! \left\| \nabla F(\boldsymbol{w}(t)) \right\|^2 
  \!\!- \! \frac{\lambda}{2I} \mathbb{E} \!\left[ \left\| \boldsymbol{\theta}(t) \right\|^2 \right] \nonumber\\
&\quad + \frac{\lambda L^2}{K} \! \sum_{k=1}^{K} \sum_{i=0}^{I - 1} \mathbb{E} \!\left[ \left\| \boldsymbol{w}(t) - \boldsymbol{w}_k(t, i) \right\|^2 \right]
+ \lambda I \xi^2.
\end{align}
\textit{Lemma 2}: Based on Assumptions~2–3 and Definition~1, the deviation between the global model and each local model admits the following upper bound:
\setlength{\jot}{1pt}
\begin{align}
\sum_{i=0}^{I - 1} 
\mathbb{E} \!\left[ \left\| \boldsymbol{w}(t) - \boldsymbol{w}_k(t, i) \right\|^2 \right]
&\le  \frac{4 \lambda^2 \xi^2 I^3}{3}
+ 4 \lambda^2 \sigma_d^2 I^3 \nonumber\\
&\quad + 4 \lambda^2 I^3 \left\| \nabla F(\boldsymbol{w}(t)) \right\|^2.
\end{align}
\textit{Lemma 3}: Suppose that Assumptions~3 and~4 hold. Then, the aggregation error $\boldsymbol{e}^T(t)=\hat{\boldsymbol{s}}^T(t) - \boldsymbol{s}^T(t)$ and the instantaneous MSE satisfy
\begin{equation}
\mathbb{E} \!\left[ \left\| \boldsymbol{e}(t) \right\|^2 \right] 
\le  \frac{\Gamma}{K^2} \text{MSE}(t).
\end{equation}
\textit{Proof of Theorem 1}: 
Given that $F(\boldsymbol{w})$ is an $L$-smooth function, we can derive the following inequality:
\setlength{\jot}{1pt}
\begin{align} \label{Theo1}
&F(\boldsymbol{w}(t + 1)) - F(\boldsymbol{w}(t)) \nonumber\\
&\le -\lambda \langle \nabla F(\boldsymbol{w}(t)), \boldsymbol{\theta}(t) + \mathbf{e}(t) \rangle
+ \frac{\lambda^2 L}{2} \| \boldsymbol{\theta}(t) + \mathbf{e}(t) \|^2 \nonumber\\
&\overset{(a)}{\le} -\lambda \langle \nabla F(\boldsymbol{w}(t)), \boldsymbol{\theta}(t) \rangle 
+ \frac{\lambda}{2} \| \nabla F(\boldsymbol{w}(t)) \|^2 
+ \frac{\lambda}{2} \| \mathbf{e}(t) \|^2 \nonumber\\
&\quad + \frac{\lambda^2 L}{2} \| \boldsymbol{\theta}(t) \|^2 
+ \frac{\lambda^2 L}{2} \| \mathbf{e}(t) \|^2 
+ \lambda^2 L \langle \boldsymbol{\theta}(t), \mathbf{e}(t) \rangle \nonumber\\
&\overset{(b)}{\le} -\lambda \langle \nabla F(\boldsymbol{w}(t)), \boldsymbol{\theta}(t) \rangle 
+ \frac{\lambda}{2} \| \nabla F(\boldsymbol{w}(t)) \|^2 \nonumber\\
&\quad + \left( \frac{\lambda}{2} + \lambda^2 L \right) \| \mathbf{e}(t) \|^2 
+ \lambda^2 L \| \boldsymbol{\theta}(t) \|^2.
\end{align}
where $(a)$ employs the inequality $-\boldsymbol{a}^T \boldsymbol{b} \le \frac{\|\boldsymbol{a}\|^2}{2} + \frac{\|\boldsymbol{b}\|^2}{2}$, and $(b)$ is derived from $\boldsymbol{a}^T \boldsymbol{b} \le \frac{\|\boldsymbol{a}\|^2}{2} + \frac{\|\boldsymbol{b}\|^2}{2}$.
Taking expectations over both sides of~\eqref{Theo1} yields
\setlength{\jot}{1pt}
\begin{align}
&\mathbb{E} \!\left[ F(\boldsymbol{w}(t+1)) - F(\boldsymbol{w}(t)) \right]
\le -\lambda \mathbb{E} \!\left[ \langle \nabla F(\boldsymbol{w}(t)), \boldsymbol{\theta}(t) \rangle \right] \nonumber\\
&\quad + \frac{\lambda}{2} \! \left\| \nabla F(\boldsymbol{w}(t)) \right\|^2
+ \left( \frac{\lambda}{2} + \lambda^2 L \right) \mathbb{E} \!\left[ \| \mathbf{e}(t) \|^2 \right] \nonumber\\
&\quad + \lambda^2 L \mathbb{E} \!\left[ \| \boldsymbol{\theta}(t) \|^2 \right].
\end{align}
By combining Lemmas~1–3 and applying the constraint $\lambda \le \min \!\left\{ \frac{1}{2LI},\, \frac{1}{\sqrt{6 L^2 I^3}},\, \frac{\sqrt{I-1}}{4LI} \right\}$, we can obtain
\setlength{\jot}{1pt}
\begin{align} \label{one_time}
&\mathbb{E} \!\left[ F(\boldsymbol{w}(t+1)) - F(\boldsymbol{w}(t)) \right]\nonumber\\
&\le -\frac{\lambda (I - 1)}{4} \! \left\| \nabla F(\boldsymbol{w}(t)) \right\|^2 
+ \left( \frac{4 \lambda^3 L^2 I^3}{3} + \lambda I \right) \xi^2 \nonumber\\
&\quad + 4 \lambda^3 L^2 \sigma_d^2 I^3
+ \left( \frac{\lambda}{2} + \lambda^2 L \right) 
N \frac{\Gamma(K+1)}{K^2} \text{MSE}(t).
\end{align}
Summing~\eqref{one_time} over all $T$ iterations gives
\setlength{\jot}{1pt}
\begin{align}
&\mathbb{E} \!\left[ F(\boldsymbol{w}(T)) - F(\boldsymbol{w}(0)) \right]
\le -\frac{\lambda (I - 1)}{4} \! \sum_{t=0}^{T-1} 
\left\| \nabla F(\boldsymbol{w}(t)) \right\|^2 \nonumber\\
&\quad + \left( \frac{4 \lambda^3 L^2 I^3}{3} + \lambda I \right) \xi^2 T 
+ 4 \lambda^3 L^2 \sigma_d^2 I^3 T \nonumber\\
&\quad + \left( \frac{\lambda}{2} + \lambda^2 L \right) 
\sum_{t=0}^{T-1} N \frac{\Gamma(K+1)}{K^2} \text{MSE}(t).
\end{align}
Finally, with Assumption~1, we have the conclusion presented in~\eqref{OBJ0} of Theorem~1.

\vspace{-0.5cm}
\section{Proof of Theorem~2}\label{Theorem2}
In this Appendix, we provide the detailed derivation steps to obtain the closed-form solution presented in Theorem~2. 
We first construct the Lagrangian function corresponding to problem~$(\mathcal{P}1.2)$, which can be expressed as
\setlength{\jot}{1pt}
\begin{align}
&\mathcal{L}\!\left(
\lbrace P_k(t)\rbrace,\lbrace \alpha_k(t)\rbrace,\lbrace \rho_k\rbrace
\right)\! = \nonumber\\
& \sum_{t=0}^{T-1}
\!\left( 
\!\left(|\boldsymbol{b}^H(t)\hat{\boldsymbol{h}}_k(t)|\sqrt{P_k(t)} - 1\right)^2 
+ P_k(t)\sigma_h^2 \|\boldsymbol{b}(t)\|^2 
\!\right) \nonumber\\
&+ \sum_{t=0}^{T-1}\!\alpha_k(t)\!\left(P_k(t)-P_k^{\max}\right)
+ \rho_k\!\left(\sum_{t=0}^{T-1}P_k(t)-TP_k^{\text{ave}}\right),
\end{align}
where $\{\alpha_k(t)\}$ and $\{\rho_k\}$ denote the dual variables corresponding to the individual and average power constraints in~\eqref{indi_power} and~\eqref{aver_power}, respectively.  

Then, we have the derivative $\partial\mathcal{L}/\partial P_k(t)$, which equals to
\setlength{\jot}{1pt}
\begin{align}
\frac{\partial \mathcal{L}}{\partial P_k(t)}
=&\frac{\left|\boldsymbol{b}^H(t)\hat{\boldsymbol{h}}_k(t)\right|}{\sqrt{P_k(t)}} 
\!\left(\left|\boldsymbol{b}^H(t)\hat{\boldsymbol{h}}_k(t)\right|\sqrt{P_k(t)}-1\right) \nonumber\\
&+ \sigma_h^2\|\boldsymbol{b}(t)\|^2 + \alpha_k(t) + \rho_k.
\end{align}
By setting $\frac{\partial \mathcal{L}}{\partial P_k(t)}=0$, the transmit power $P_k(t)$ satisfies
\begin{equation}
P_k(t)
=\left(
\frac{\left|\boldsymbol{b}^H(t)\hat{\boldsymbol{h}}_k(t)\right|}
{\left|\boldsymbol{b}^H(t)\hat{\boldsymbol{h}}_k(t)\right|^2
+\sigma_h^2\|\boldsymbol{b}(t)\|^2+\alpha_k(t)+\rho_k}
\right)^{\!2}.
\end{equation}

Let $P_k^*(t)$ denote the optimal transmit power for problem $(\mathcal{P}1.1)$, $\alpha_k^*(t)$ and $\rho_k^*$ represent the associated optimal dual variables. 
The corresponding KKT conditions can be expressed as:
\setlength{\jot}{1pt}
\begin{align}
&P_k^*(t)
=\!\left(
\frac{|\boldsymbol{b}^H(t)\hat{\boldsymbol{h}}_k(t)|}
{|\boldsymbol{b}^H(t)\hat{\boldsymbol{h}}_k(t)|^2
+\sigma_h^2\|\boldsymbol{b}(t)\|^2+\alpha_k^*(t)+\rho_k^*}
\right)^{\!2}, \label{stationary}\\
&0 \le P_k^*(t) \le P_k^{\max}, \label{Max_power}\\
&0 \le \sum_{t=0}^{T-1}P_k^*(t)\le T P_k^{\text{ave}}, \label{ave_power}\\
&\alpha_k^*(t)\ge0,\quad \rho_k^*\ge0,\\
&\alpha_k^*(t)\!\left(P_k^*(t)-P_k^{\max}\right)=0, \label{sla1}\\
&\rho_k^*\!\left(\sum_{t=0}^{T-1}P_k^*(t)-TP_k^{\text{ave}}\right)=0. \label{sla2}
\end{align}

\subsection{Maximum Transmit Power Case}
To analyze the maximum transmit power condition in~\eqref{sla1}, two scenarios can be identified:

1) When $\alpha_k^*(t)>0$, the complementary slackness condition in~\eqref{sla1} implies $P_k^*(t)=P_k^{\max}$;

2) When $\alpha_k^*(t)=0$, we obtain
\begin{equation}
P_k^*(t)=
\left(
\frac{|\boldsymbol{b}^H(t)\hat{\boldsymbol{h}}_k(t)|}
{|\boldsymbol{b}^H(t)\hat{\boldsymbol{h}}_k(t)|^2
+\sigma_h^2\|\boldsymbol{b}(t)\|^2+\rho_k^*}
\right)^{\!2}.
\end{equation}
Therefore, the optimal transmit power can be compactly written as
\begin{equation}\label{Final_Sla1}
P_k^*(t)=
\min\!\left\{
\!\!\left(
\frac{|\boldsymbol{b}^H(t)\hat{\boldsymbol{h}}_k(t)|}
{|\boldsymbol{b}^H(t)\hat{\boldsymbol{h}}_k(t)|^2
+\sigma_h^2\|\boldsymbol{b}(t)\|^2+\rho_k^*}
\right)^{\!2},
P_k^{\max}
\!\!\right\}.
\end{equation}

From~\eqref{Final_Sla1}, it is observed that $P_k^*(t)$ is a function of $\rho_k^*$. 
Next, we examine how the average transmit power constraint influences the optimal power allocation across local devices.

\subsection{Average Transmit Power Case}
We now consider the average transmit power constraint in~\eqref{sla2}. Two cases arise:

1) If $\rho_k^*>0$, \eqref{sla2} gives $
\sum_{t=0}^{T-1} P_k^*(t)=T P_k^{\text{ave}}$, which leads to \eqref{Final_Sla1},
with $\rho_k^*$ determined by bisection to meet the constraint.

2) When $\rho_k^*=0$, if $\sum_{t=0}^{T-1}P_k^*(t)\le T P_k^{\text{ave}}$, we have
\begin{equation}\label{Final_Sla2}
P_k^*(t)=
\min\!\left\{
\left(
\frac{|\boldsymbol{b}^H(t)\hat{\boldsymbol{h}}_k(t)|}
{|\boldsymbol{b}^H(t)\hat{\boldsymbol{h}}_k(t)|^2
+\sigma_h^2\|\boldsymbol{b}(t)\|^2}
\right)^{\!2},
P_k^{\max}
\right\}.
\end{equation}
Otherwise, if $\rho_k^*=0$ and $\sum_{t=0}^{T-1}P_k^*(t)>T P_k^{\text{ave}}$, the constraint in~\eqref{ave_power} is violated, implying that $\rho_k^*>0$ must hold, which leads to the result in~\eqref{Final_Sla1}. 
Combining the two cases yields the expression stated in Theorem~2.

\section{Proof of Theorem 3} \label{Theorem3}
The objective function can be expressed as 
\setlength{\jot}{1pt}
\begin{align}
&\mathcal{F}(\boldsymbol{b}(t)) = \sum_{k=1}^{K} 
\left| \mathbf{b}^H(t) \hat{\mathbf{h}}_k(t) \mu_k(t) - 1 \right|^2 \nonumber
\\&\quad\quad\quad\quad\quad\quad\quad+ \left( \sum_{k=1}^{K} P_k(t) \sigma_h^2
+ \sigma_0^2 \right) \|\boldsymbol{b}(t)\|^2.
\end{align}
let $\beta=\left( \sum_{k=1}^{K} P_k(t) \sigma_h^2
+ \sigma_0^2 \right)$, we have 
\setlength{\jot}{1pt}
\begin{align}
&\mathcal{F}(\boldsymbol{b}(t)) = \sum_{k=1}^{K} \left| \boldsymbol{b}^H(t)\hat{\boldsymbol{h}}_k (t)\mu_k(t) - 1 \right|^2 
+ \beta \|\boldsymbol{b}(t)\|^2,
\end{align}
which is convex and has a unique global minimum. To obtain the solution,
we first expend the first term as
\setlength{\jot}{1pt}
\begin{align}
    &\left| \boldsymbol{b}^H(t)\boldsymbol{\delta}_k(t) - 1 \right|^2= \nonumber\\&\boldsymbol{b}^H(t)\boldsymbol{\delta}_k (t)\boldsymbol{\delta}_k^H (t) \boldsymbol{b}(t)-\boldsymbol{b}^H(t)\boldsymbol{\delta}_k(t)-
\boldsymbol{\delta}_k^H(t)\boldsymbol{b}(t)+1,
\end{align}
where $\boldsymbol{\delta}_k(t)=\hat{\boldsymbol{h}}_k (t)\mu_k(t)\in \mathbb{C}^{M \times 1}$. Thus $\mathcal{F}(\boldsymbol{b}(t))$ becomes
\setlength{\jot}{1pt}
\begin{align}
\mathcal{F}(\boldsymbol{b}(t))= &  \boldsymbol{b}^H(t)\left(
\sum_{k=1}^K\boldsymbol{\delta}_k (t)\boldsymbol{\delta}_k^H (t)+\beta\mathbf{I}_M
\right)\boldsymbol{b}(t)\nonumber\\
&-2Re\left\lbrace\sum_{k=1}^K\boldsymbol{b}^H(t)\boldsymbol{\delta}_k(t) \right\rbrace+K. 
\end{align}
The first order derivative of $\mathcal{F}(\boldsymbol{b}(t))$ can be expressed as
\begin{equation}
    \nabla\mathcal{F}(\boldsymbol{b}(t))=2\left(
\sum_{k=1}^K\boldsymbol{\delta}_k (t)\boldsymbol{\delta}_k^H (t)+\beta\mathbf{I}_M
\right)\boldsymbol{b}(t)-2\sum_{k=1}^K\boldsymbol{\delta}_k(t),
\end{equation}
but setting $\nabla\mathcal{F}(\boldsymbol{b}(t))=0$, we have the unique optimal solution for 
$(\mathcal{P}2.1)$ as
\begin{equation}
   \boldsymbol{b}^*(t) =\left(
\sum_{k=1}^K\boldsymbol{\delta}_k (t)\boldsymbol{\delta}_k^H (t)+\beta\mathbf{I}_M
\right)^{-1}\sum_{k=1}^K\boldsymbol{\delta}_k(t),
\end{equation}
\endgroup
which leads to Theorem 3.

\bibliographystyle{IEEEtran}
\footnotesize

\bibliography{Misalignment_Ref}
\vspace{-3em}
\begin{IEEEbiographynophoto}{Xiaoyan Ma}  (Member, IEEE) received the B.S. and Ph.D. degrees from the School of Information Science and Engineering, Shandong University, Jinan,
China, in 2017 and 2023, respectively. She is currently pursuing the second Ph.D. degree with Purdue University, West Lafayette, IN, USA. From
2023 to 2024, she was a Research Fellow with the School of Electrical and Electronic Engineering,
Nanyang Technological University, Singapore. Her research interests include multiple-in–multiple-out
technology, signal processing and federated learning.
\end{IEEEbiographynophoto}
\vspace{-3em}
\begin{IEEEbiographynophoto}{Shahryar Zehtabi} (Graduate Student Member,
IEEE) received the B.Sc. degrees in EE and CE from the Amirkabir University of Technology in 2020 and 2021, respectively, and the M.Sc. degree in ECE from Purdue University in 2024, where he is currently pursuing the Ph.D. degree in ECE. He was a recipient of the 2024 and 2026 Magoon Award for Excellence in Teaching and Excellence in Research, respectively.
\end{IEEEbiographynophoto}
\vspace{-3em}
\begin{IEEEbiographynophoto}{Yinan Zou}  (Student Member, IEEE) is currently pursuing the Ph.D. degree with Purdue University.
\end{IEEEbiographynophoto}
\vspace{-3em}
\begin{IEEEbiographynophoto}{Taejoon Kim} (Senior Member, IEEE) received the
Ph.D. degree in electrical and computer engineering from Purdue University, West Lafayette, IN, USA.
He is currently an Associate Professor with the School of Electrical, Computer and Energy Engineering, Arizona State University (ASU). Before
joining ASU, he held positions with Nokia, KTH Royal Institute of Technology, City University of
Hong Kong, and University of Kansas (KU). He serves as an Associate Editor for IEEE TRANSACTIONS ON WIRELESS COMMUNICATIONS and IEEE OPEN JOURNAL OF THE COMMUNICATION SOCIETY. Previously, he served as an Associate Editor of IEEE TRANSACTIONS ON COMMUNICATIONS and a Guest Editor for IEEE TRANSACTIONS ON INDUSTRIAL
INFORMATICS. His research spans communications, signal processing, and security. He was a recipient of the Harry Talley Excellence in Teaching Award, Kansas Board of Regent Faculty of the Year Award, and the Miller Professional Development Award in Research. Along with the co-authors, he won the IEEE Communications Society Stephen O. Rice Prize in 2016 and the IEEE PIMRC 2012 Best Paper Award.
\end{IEEEbiographynophoto}
\vspace{-3em}
\begin{IEEEbiographynophoto}{Christopher G. Brinton} (Senior Member, IEEE) received the M.S. and Ph.D. (Hons.) degrees in electrical engineering from Princeton University in 2013 and 2016, respectively. He is currently a Elmore Associate Professor of electrical and computer engineering (ECE) and the Faculty Director of the Hub for Innovation, Ventures, and Entrepreneurship (HIVE), Purdue University. Prior to joining Purdue, he was the Associate Director of the EDGE Laboratory and a Lecturer of electrical engineering with Princeton University. His research interests include the intersection of machine learning, communications, and networking, specifically in edge intelligence, AI/ML-inspired wireless design, and foundation models. He was a recipient of five of U.S. top early career awards from the National Science Foundation (CAREER), the Office of Naval Research (YIP), the Defense Advanced Research Projects Agency (YFA and Director’s Fellowship), and the Air Force Office of Scientific Research (YIP). He was also a recipient of the IEEE Communication Society William Bennett Prize Best Paper Award, the Intel Rising Star Faculty Award, the Qualcomm Faculty Award, several conference paper awards, and the Purdue College of Engineering (CoE) Faculty Excellence Award in Early Career Research.
\end{IEEEbiographynophoto}

\end{document}